\documentclass{aa}
\usepackage{graphicx}
\usepackage{xcolor}
\usepackage{txfonts}
\usepackage{natbib,twoopt}

\usepackage[breaklinks,unicode=true,colorlinks,linkcolor=red,citecolor=red,pdfdisplaydoctitle]{hyperref} 
\bibpunct{(}{)}{;}{a}{}{,}             
\makeatletter
  \newcommandtwoopt{\citeads}[3][][]{\href{http://adsabs.harvard.edu/abs/#3}%
    {\def\hyper@linkstart##1##2{}%
     \let\hyper@linkend\@empty\citealp[#1][#2]{#3}}}
  \newcommandtwoopt{\citepads}[3][][]{\href{http://adsabs.harvard.edu/abs/#3}%
    {\def\hyper@linkstart##1##2{}%
     \let\hyper@linkend\@empty\citep[#1][#2]{#3}}}
  \newcommandtwoopt{\citetads}[3][][]{\href{http://adsabs.harvard.edu/abs/#3}%
    {\def\hyper@linkstart##1##2{}%
     \let\hyper@linkend\@empty\citet[#1][#2]{#3}}}
  \newcommandtwoopt{\citeyearads}[3][][]%
    {\href{http://adsabs.harvard.edu/abs/#3}
    {\def\hyper@linkstart##1##2{}%
     \let\hyper@linkend\@empty\citeyear[#1][#2]{#3}}}
\makeatother
\begin{document} 

   \title{Semi-automatic meteoroid fragmentation modeling using genetic algorithms}
   \titlerunning{Meteoroid fragmentation modeling using genetic algorithms}
   \author{Tom\'a\v{s} Henych\inst{1}\fnmsep\thanks{Corresponding author, email: ftom@physics.muni.cz}
     \and
     Ji\v{r}\'i Borovi\v{c}ka\inst{1}
     \and
     Pavel Spurn\'{y}\inst{1}
   }
   \institute{Astronomical Institute, Czech Academy of Sciences, 251 65 Ond\v{r}ejov, Czech Republic}
   \date{Received 20 September, 2022; accepted 19 January 2023}

 
  \abstract
    {Meteoroids are pieces of asteroids and comets. They serve as unique 
probes to the physical and chemical properties of their parent bodies. 
We can derive some of these properties when meteoroids collide with the 
atmosphere of Earth and become a meteor or a bolide. Even more 
information can be obtained when meteoroids are mechanically strong and 
slow enough to drop meteorites.}
   {Through physical modeling of bright meteors, we describe their 
fragmentation in the atmosphere. We also derive their mechanical strength 
and the mass distribution of the fragments, some of which may hit the 
ground as meteorites.}
   {We developed a semi-automatic program for meteoroid fragmentation 
modeling using parallel genetic algorithms. This allowed us to determine 
the most probable fragmentation cascade of the meteoroid, and also to 
specify its initial mass and velocity. These parameters can be used in turn 
to derive the heliocentric orbit of the meteoroid and to place constraints 
on its likely age as a separate object.}
   {The program offers plausible solutions for the majority of fireballs 
we tested, and the quality of the solutions is comparable to that of 
manual solutions. The two solutions are not the same in detail, but the 
derived quantities, such as the fragment masses of the larger fragments 
and the proxy for their mechanical strength, are very similar. With this 
method, we would like to describe the mechanical properties and 
structure of both meteoroids belonging to major meteor showers and those 
that cause exceptional fireballs.}
   {}

   \maketitle

\section{Introduction}
Meteoroids are fragments of comets and asteroids. They are therefore a 
valuable source of information about those objects. Observations 
of meteoroids can provide us with clues about their structure, density, and mechanical 
strength. If they produce meteorites, we can also derive their porosity, 
mineralogy, and chemical composition, and all sorts of other physical 
properties. Meteorite-producing fireballs are also a very rare opportunity 
for a cheap sampling of material from asteroids, which greatly increases 
the scientific return of their observations.

Research of asteroids and comets brings a wealth of information about 
the physical and chemical properties of the protosolar nebula. These bodies 
also provide constraints on the process of formation and the following 
evolution of the Solar System. Moreover, revealing the history of our 
Solar System enables important insights into the formation and evolution 
of other planetary systems, including the formation of planets similar to 
our Earth that can potentially host life.

Moreover, even small asteroids pose a threat to life on Earth, as was 
demonstrated, for example, in February 2013 by the Chelyabinsk meteorite fall 
\citep{borovicka2013a,brown2013,popova2013}. This event caused many 
injuries, and more than 1600 people sought medical help in hospitals 
\citep{akimov2015,kartashova2018}. It served as a warning sign and a 
reminder that asteroid and fireball observations are also important in 
the context of planetary defense. These observations allow us to 
calculate the likelihood of being hit by a projectile of a 
certain size more precisely, and they enable us to estimate the possible outcomes of such a 
collision for our civilization as well as the environment.

Many characteristics of a meteoroid can easily be derived from the
fireball observations, but some parameters are
dependent on detailed physical modeling and comparison with all the
available observations. Classical models include those reported by \citet{baldwin1971} and 
\citet{hills1993}. Hydrodynamical modeling of outcomes of the comet 
Shoemaker-Levy~9 impact on Jupiter was reported by \citet{ahrens1994} and 
\citet{boslough1994}. The disintegration of larger meteoroids and 
asteroids in the atmosphere of Earth was modeled by \citet{svetsov1995}. 
A more recent fragment-cloud model of the asteroid atmospheric breakup 
and energy deposition was introduced in \citet{wheeler2017} and was 
extended and evaluated for various internal structures of meteorids by 
\citet{wheeler2018}. A recent detailed review of various approaches to 
meteoroid entry modeling is given in \citet{popova2019}.

Our paper extends the modeling efforts of \citet{borovicka2013b} ,
who successfully modeled the Košice meteorite fall with their 
semi-empirical model. They were not only able to predict very precisely where 
the meteorites would be found, but also described the fireball phenomenon 
in great detail. That modeling also included the effect of the so-called 
erosion, which was previously used for the successful modeling of Draconid 
meteors \citep{borovicka2007}. Further improvements of the model are 
reflected in the recent paper of \citet{borovicka2020}. 

The main goal of these modeling efforts is to find a precise initial mass 
and velocity of the meteoroid, fragmentation times that enable us to 
derive the bulk strength of a fragmenting body, the number and physical 
properties of daughter fragments that originated in the meteoroid 
fragmentation cascade, and the mass distribution of the dust particles. 
Moreover, in some exceptional cases, we would like to derive the masses 
and distribution of potential meteorites after additional modeling of 
the dark flight. In other words, we are interested in the internal 
structure and also in the mass and strength distributions of the meteoroid and 
its constituents.

The trial-and-error method of \citet{borovicka2020} was used for manual 
modeling with the Firmodel program. It produces a good eyeball fit of 
the radiometric curve, the photometric light curve, and the dynamics of 
the fireball.

The manual approach, however, is time-consuming, and it searches the
parametric space of the problem randomly. The solution may not be 
unique, and some better solutions that are not found in this 
way may even exist. Therefore, we develop a semi-automatic modeling 
procedure that searches the parametric space more thoroughly and can find more solutions that can then be compared. The 
best solution can finally be chosen.

The European Fireball Network (EN) has provided observations of fireballs
over Central Europe for decades. It is being actively developed and
expanded by new stations and new instrumentation \citep{spurny2007,
spurny2017}. Because the sky is monitored for bright fireballs every 
night, and because the observing methods are gradually improved and even expanded, 
the number of high-quality observations is ever-growing. The processing 
of the raw data is now well established, and the only remaining step 
toward deriving meteoroid physical characteristics is data modeling.

When developing the semi-automatic modeling procedure, our first idea 
was to search the parametric space as thoroughly as possible. However, 
because the problem is so complex, the parametric space is vast. No traditional method would complete the search in a 
reasonable time, and/or a traditional method would probably only find some local optimum. 
Therefore, we envisioned a semi-automatic approach to the problem by the 
use of parallel genetic algorithm (GA) optimization. In this way, 
it should be possible to find a solution that optimizes all data of the fireball (radiometric and photometric light curves, and 
dynamics), and this solution may also be similar to a manual solution of the 
problem. \citet{tarano2019} were the first who used the 
combination of a GA optimizer with a meteoroid 
fragmentation model, namely the fragment-cloud model of 
\citet{wheeler2017}, to derive the physical characteristics of meteoroids. 
Later, \citet{tarano2020} expanded 
this work by using supervised learning methods and suggested further 
improvements of the meteoroid characteristics inference.

Compared to \citet{tarano2019}, there are several important differences. 
First, our luminous efficiency is not a constant, but a function that 
depends on both the mass and the velocity of the meteoroid. Second, we 
can calculate the meteoroid initial velocity and height from the 
dynamics data alone, while the initial (photometric) mass is estimated 
from the radiometric curve. All of these values are then optimized while 
 the solution is found, but the initial estimates dramatically shrink the 
search space. Third, we optimize three separate datasets: radiometric 
and photometric light curves, and dynamics. These data carry more detailed 
information about the meteoroid entry than the energy deposition curve 
that was used in \citet{tarano2019} and other studies. Fourth, we use a 
different fragmentation model that is more driven by the data itself, 
that is, fragmentation times are deduced from the radiometric curve and 
are verified by the dynamics. Further details can be found below in 
Sect.~\ref{firmpik-program}.

The algorithm extensively uses pseudo-random numbers~at the
initialization and during the calculation~to choose the values of 
the free parameters. From the previous use of GA, we know
that it can converge to a global optimum of the problem.
Therefore, the algorithm is probably capable of finding a unique
set of solutions for the fireball fragmentation in the atmosphere as well.
We expect that there will be several solutions to the problem with a
similar quality that may differ in details or differ even more
substantially. When we obtain these solutions, we can compare them and
determine the best of them, or we can define what a general solution looks like.
These solutions may differ in the particular number of fragments, their
respective masses, and other parameters. Nevertheless, we should be able
to find a plausible range of these values. If we succeed in finding the
global optimum of the problem, we can then calculate the uncertainties
of the fit by inspecting its close vicinity by a systematic search or by
a Monte Carlo method.

Section~\ref{data-modeling} contains the description of observation methods 
and data processing, and Sect.~\ref{firmpik-program} introduces FirMpik,~the semi-automatic program for data modeling. In Sect.~\ref{results}, and in 
Appendix~\ref{app_fig} we summarize our results, which we discuss in 
Sect.~\ref{discussion}, and Sect.~\ref{conclusions} lists our main 
conclusions and future plans.

\section{Data and their modeling}\label{data-modeling}

\subsection{Observations}\label{observations}
All the modeled data come from the European fireball network, which
comprises 21 stations in central Europe: 15 stations are in the Czech
Republic, 4 in Slovakia, and one each in Austria and in Germany. These 
stations are 
equipped with the Digital Autonomous Fireball Observatory, or DAFO, 
which has been developed at the Astronomical Institute of The Czech 
Academy of Sciences in Ond\v{r}ejov. The observatory contains two digital 
all-sky cameras and a sensitive photomultiplier tube pointed toward the 
zenith that produces very precise high-time-resolution radiometric data. 
Some stations also contain narrow-field cameras for detailed 
observations of meteoroid fragmentation. The fireball trail in the 
all-sky image is interrupted by an LCD shutter behind the lens with a 
frequency of $16\,\rm{Hz}$ (this can be modified remotely) to enable 
a measurement of the fireball velocity. For temporal 
calibration, one interruption is omitted every whole second, producing a 
longer dash on the fireball trail. Time at respective stations is 
continuously corrected by the pulse-per-second (PPS) signal of the GPS ,
providing an absolute timing error better than $1\,\rm{\mu s}$. Because all 
the data are digital, they are daily 
uploaded to a central server and are immediately available. The data are
preprocessed, and bright fireballs are automatically detected, which 
enables a timely modeling.

\subsection{Data reduction}\label{data-reduction}
The data on the brightest fireballs are reduced on the day after
the observations. The radiometric curve is calibrated to absolute
magnitudes (fireball brightness at a $100\,\rm{km}$ distance from the station)
by comparison with digital photographs from all-sky cameras. The all-sky
image with a captured fireball is calibrated both astrometrically and
photometrically, and then the positions and brightness of individual
parts of the trail are measured. These measurements with time
calibrations produce a light curve that provides valuable information 
of the fainter parts of the fireball (the beginning and the end) because 
the radiometer is less sensitive than the digital camera. On the other 
hand, the radiometer has a much wider dynamic range than the camera.
Fireball observations from two or more stations enable us to calculate 
its trajectory in the atmosphere. The dynamics of the foremost meteoroid 
fragment, or its length along the trajectory versus time, is calculated 
from the position measurements of individual breaks of the fireball 
trail caused by the LCD shutter \citep{borovicka2020}.

\subsection{Modeling}\label{modeling}
In the original version of the Firmodel program, the modeling is performed in
a trial-and-error fashion. The user fits the data by eye, in other
words, there is no objective measure of the fit in the process, although
the fit is checked in the O$-$C plot. This plot shows the 
difference between the observed (O) and calculated (C) values of the 
length along the trajectory of the fireball (length residuals) or the 
fireball brightness measured from the all-sky images or by a radiometer 
as a function of time or height above the ground. The automatic 
algorithm we developed also uses a trial-and-error method together with 
simplified but very powerful rules of natural selection, inheritance, and 
variation. As a fitness function (an objective function specific for 
GA, opposite of the cost function) that measures the 
quality of each solution, we use the inverse value of the reduced $\chi^2$ 
sum of the model fit to all available data. The reduced $\chi^2$ is 
defined as

\begin{equation}
\chi^2=\frac{w_{\rm rlc}\chi^2_{\rm rlc}+w_{\rm lc}\chi^2_{\rm lc}+w_{\rm len}\chi^2_{\rm len}}{N_{\rm data}-N_{\rm parm}}
,\end{equation}

where $w_{\rm rlc}$, $w_{\rm lc}$ , and $w_{\rm len}$ denote the weights of 
the three datasets we use in finding the fragmentation model (more 
details are given in Sect.~\ref{dataset-weights}), $N_{\rm data}$ 
is the total number of datapoints, and $N_{\rm parm}$ is the number of 
free parameters sought in the modeling. The respective contributions to 
the total $\chi^2$ sum from the three datasets are calculated as

\begin{equation}
\chi^2_{\rm rlc}=\sum w_{\rm frag}(rlc_{\rm smooth}-M_{\rm model})^2/\sigma^2_{\rm rlc},
\end{equation}

\begin{equation}
\chi^2_{\rm lc}=\sum(lc-M_{\rm model})^2/\sigma^2_{\rm lc},
\end{equation}

\begin{equation}
\chi^2_{\rm len}=\sum (len-L_{\rm model})^2.
\end{equation}

Here $w_{\rm frag}$ refers to the weights used to emphasize the importance of 
fragmentations in the radiometric curve (see details in Sect.~\ref{rlc-weights}), $rlc_{\rm smooth}$ is the measured brightness from the 
median-smoothed radiometric curve, $\sigma_{\rm rlc}$ is the intrinsic 
uncertainty of the brightness measured by the radiometer, 
$M_{\rm model}$ is the total brightness calculated from the model, $lc$ 
is the fireball brightness measured from the all-sky image for 
individual breaks, $\sigma_{\rm lc}$ is the brightness intrinsic 
uncertainty, $len$ is the measured length of the leading fragment, and 
$L_{\rm model}$ is its calculated length. No uncertainty is related 
to the measured length, and it is therefore not part of our calculations. 
The datapoint uncertainties (more precisely, $1/\sigma^2$) serve as 
weights to emphasize the effect of datapoints with a lower uncertainty on 
the overall $\chi^2$ sum and reduce the effect of datapoints with a higher 
uncertainty. The weights for length datapoints are all set to one.

All the sum symbols imply summing over all datapoints in the respective 
datasets. Because the model is calculated with a finite time resolution, 
the values used to compute the $\chi^2$ sum are linearly 
interpolated for the exact time instants of observations. Because the 
problem we model is highly nonlinear and also because of the 
weights used in the calculation, the $\chi^2$ does not have a strict 
statistical meaning, but rather serves as a relative measure of the 
goodness of fit. This also has implications for the confidence intervals of 
the quantities found by the modeling. In a typical optimization run, the 
value of the fitness function grows very fast at the beginning, curves over later in the process, and reaches a plateau with an occasional 
sudden rise.

The data for the radiometric curve are very precise (except for the beginning 
and end of the curve, because the signal-to-noise ratio is low), and the data from different observing 
stations are clearly related. Still, we observe some spread in the brightness values that are due to 
varying properties of radiometers and different weather conditions. 
Therefore, the user who 
determines the model manually can use this spread to hide some minor 
fragmentations that optimize both the radiometric data and dynamics. In contrast, the automatic fitting algorithm uses an objective measure (the reduced $\chi^2$ sum with respect to the smoothed radiometric 
curve, light curve, and dynamic data) and any departure from these data 
is penalized by an increase in the $\chi^2$ sum. It is therefore more 
critical to find the fragmentations in the radiometric curve as reliably 
as possible so that the fit can optimize all the available data.

The user who fits the fireball data manually also has more
information about the data quality. The radiometric data have intrinsic
uncertainties that are used for the $\chi^2$ sum calculation. The
uncertainties reflect objective observational conditions at the
respective station, but some problem
with the data may occasionally be hidden that does not propagate to its uncertainties. The user can
check the suspect data (which are poorly fit by the model) and
decide to ignore them or give them a lower priority while constructing
the model. The automatic algorithm, however, does not have this
additional information. Moreover, the dynamic data have no a priori 
uncertainties whatsoever. The model also assumes a steady-state ablation, 
which is not satisfied at the beginning of the observation when the 
meteoroid is gradually heated up and only starts to shine. The 
brightness of the fireball as predicted by the model is then higher 
than observed. This means that even though the automatic algorithm computes the 
reduced $\chi^2$ value for every solution, it mainly serves as a 
relative measure of the fit quality. A solution is accepted when the 
overall quality of the fit to all the datasets by a model is good enough 
as judged by the user, not only by the $\chi^2$ sum.

\section{FirMpik program}\label{firmpik-program}
FirMpik is a global optimization program for semi-automatic modeling
of fireball fragmentation in the atmosphere. It uses GA 
to search for several tens of parameters describing the ablation,
radiation, and fragmentation of a meteoroid during the visible flight.
These parameters include the number of fragments released from their
respective parents in gross fragmentation events, their
respective masses, whether or not the fragment erodes, and if it does, the erosion coefficient, and the mass distribution limits of
dust in both the gross and continuous fragmentations. The
model is created for the whole observation dataset at once, and it is
compared by means of reduced $\chi^2$ statistics to all the data that are
available for a specific fireball observation. These data are comprised
of the radiometric light curve taken by a sensitive photomultiplier tube
pointed to the zenith, optical light curve and dynamic data measured
from images taken by digital all-sky cameras, and occasionally, also
individual fragment observations from narrow-field cameras placed on
several stations of the Czech part of the EN. FirMpik uses the same 
fragmentation model as the Firmodel program.

\subsection{Firmodel program}\label{firmodel-program}
Firmodel is a computer program developed by Ji\v{r}\'i Borovi\v{c}ka 
that is used for the manual modeling of atmospheric fragmentation of 
meteoroids. It was first used to model the Ko\v{s}ice meteorite fall 
\citep{borovicka2013b}. The model is described in great detail in 
\citet{borovicka2020}; we describe the model briefly below.

The fragmentation model used in the Firmodel program is semi-empirical
in the sense that the fragmentation times have to be determined in the
modeling process from the observed data of the modeled fireball. These
comprise a radiometric curve, a photometric light curve, and dynamics. The 
radiometric curve of most fireballs contains very fast semi-periodic 
brightness changes. They might be caused by an instability of ablation, 
but the present model does not account for these brightness 
changes. Therefore, in modeling the bolide radiation, we use a smoothed 
radiometric curve rather than the original data.

The inputs are the initial meteoroid height, the velocity deduced from the
dynamics in the first half of its trajectory, the initial mass derived from
the total radiated energy, the calculated trajectory, and constant 
parameters describing the meteoroid bulk density ($\delta$), shape 
($A$), and also drag ($\Gamma$) and ablation ($\sigma$) coefficients. 
For the bulk density, we assumed a constant value of 
$3500\,\rm{kg\cdot m^{-3}}$, the product 
$\Gamma A=0.8$ and $\sigma=0.005\,\rm{kg\cdot MJ^{-1}}$.
These values were also inherited by all the fragments and dust released
from the meteoroid to keep the number of free parameters reasonably low. 
The initial height, velocity, and mass were then adjusted in the 
modeling process. The luminous efficiency used in the model depends on the mass and the velocity of the meteoroid, and it is the same 
function as in \citet{borovicka2020}. For our calculations, we used 
densities from the NRLMSISE-00 atmosphere model \citep{picone2002}.

A finite number of fragments is used in the model. The fragments move and
ablate independently. The motion, ablation, and radiation of fragments
are calculated according to the basic physical theory of meteors
\citep{ceplecha1998}. The closed-form integral solution of the drag and ablation
equations of \citet{ceplecha1998} yields values of the position, velocity, 
mass, and luminosity as a function of time. These equations are solved 
between fragmentation points for each fragment until it breaks apart or 
until its velocity decreases below $2.5\,\rm{km\cdot s^{-1}}$ , when its 
ablation (and radiation) ends. The dynamics of the foremost fragment is 
measured from the data and compared with the model, while the luminosity 
from the radiometric curve is compared with the sum of the respective
luminosities of individual fragments and dust particles in the model.

The code includes several types of fragments. Individual fragments
form in gross fragmentation events and can fragment repeatedly.
Multiple fragments are identical fragments created to save computational
time. It is an idealization in the model when the computation is
performed only once and the resulting luminosity is multiplied by the
number of such multiple fragments. Dust is released either suddenly in
gross fragmentation or by the so-called erosion, which is described in
Sect.~\ref{erosion}. Its mass distribution is a power-law function with 
two mass limits and a slope. The sudden release 
of small dust particles produces a short bright flare, while larger dust 
particles cause a longer and asymmetric peak with a quick rise and 
slower decay in brightness. Dust particles can no longer fragment, they 
are subjected to ablation only. The formation of multiple large 
fragments produces a long-lasting sudden brightening or step observed in 
the radiometric curve.

\subsection{Erosion}\label{erosion}
The meteoroid can break up suddenly, which can be demonstrated as a bright
flare in the radiometric curve if dust is also released, or continually
 (this fragmentation is 
called erosion), which is manifested by a gradual brightening or a hump in 
the radiometric curve. This phenomenon is based on the concept of a 
dustball meteoroid \citep{opik1955}. It was first used to model the light curves of Draconid fireballs in \citet{borovicka2007}, where a 
rigorous description can be found.

In such a fragmentation event, individual grains are released from a
parent fragment. The grains then behave as individual independent meteors.
A power-law function models their mass distribution, that is, four
parameters describe the distribution, the total mass released from the
parent fragment, two mass limits, and a power-law index. Nevertheless, it is often sufficient to use a single mass for the released particles. 
Moreover, an analogous parameter to the ablation coefficient 
is used to describe erosion. It is called the erosion coefficient, $\eta$,
and the erosion rate is then

\begin{equation}
\left(\frac{{\rm d} m}{{\rm d} t}\right)_{\rm{erosion}}=-K\eta m^{2/3}\rho v^3,
\end{equation} 

where $m$ is the meteoroid mass, $v$ is its velocity, $\rho$ is the 
density of the atmosphere, and $K$ is the shape-density coefficient, defined as

\begin{equation}
K=\Gamma A \delta^{-2/3}.
\end{equation} 

To further limit the number of free parameters, we only chose the erosion
coefficient and one grain mass limit for both ends of the distribution,
that is, only one value of the grain mass. The erosion coefficient was 
chosen from 20 discrete values in the range of 
$(0.05\rm{-}5)\,\rm{kg\cdot MJ^{-1}}$ , and the condition must hold that 
$\eta>5\sigma$ (when some general value of $\sigma$ is used). The 
decimal logarithm of the grain mass was also picked from discrete values 
with a step of $0.5$. It is higher than $-11$ and lower by $2.5$ than the 
decimal logarithm of mass of the eroding fragment from which the grains 
are released. For detailed modeling, we freed the two mass limits and set 
the power-law mass distribution index to two. The eroding fragment in 
our model cannot undergo another fragmentation.

The program was recently updated to run faster and more efficiently. It
allows calculating the erosion in longer time steps with almost the same
precision, but with a substantial speedup of the computation.

\subsection{Global optimization}\label{global-optimization}
As indicated in the previous section, the parametric space of the
problem is vast and high dimensional and probably contains many
local minima. To be able to find a good solution to the
problem, that is, a global minimum, we need to use a robust global
optimization algorithm. We decided to use GA, which are
adaptive heuristic search algorithms and a part of the evolutionary
algorithms group. These algorithms have been proven to be able to solve
hard optimization problems in various scientific and engineering fields.
\citet{tarano2019} used GA to infer the 
physical characteristics of meteoroids  and were able to derive similar values of the 
initial mass, initial strength, and ablation coefficient for three 
well-described fireballs as other authors in preceding papers. In 
other astrophysics areas, successful use has been demonstrated by \citet{charb1995}, for instance. In the next section, we give a brief
overview of the principles of GA.

\subsection{Genetic algorithms}\label{genetic-algorithms}
GA are a class of evolutionary algorithms or search
techniques that are inspired by the natural selection described by Charles
Darwin \citep{darwin1859}. They use three simplified forms of the 
biological processes or rules observed in nature: selection, inheritance, 
and variation. The first rule is implemented by judging the quality of the
individual solutions by the value of the so-called fitness function. The
second rule means that parts of the best solutions are passed on through
generations, and the third rule is realized by a random mutation or a
disturbance of the solution.

The fitness function is the key parameter that is used to quantify the
quality of the solution for the purpose of evolution. Usually, an
inverse of the $\chi^2$ value is used as a fitness function to compare
the model to the observed data. We used this approach, but there
are other possibilities to judge the quality of the solution.

Another important part of the GA is the representation of
trial solutions and the application of the three rules given above. As
in genetics, the algorithm works with a phenotype and a genotype. The 
first term represents the free parameters of the problem we try 
to optimize, they are the visible traits. In our case, the free 
parameters comprise the initial mass, velocity, and height 
above the ground of the meteoroid, the number of daughter fragments released in a specific 
fragmentation time, their masses, whether or not the daughter fragment 
is eroding, and if it does, the erosion coefficient $\eta$, the 
mass limit for the eroded debris, and finally, the mass limit for the dust 
released from the parent fragment. The procedure of choosing these 
parameters is further described in Sect.~\ref{optimization-function}.

The genotype designates some simplified string that encodes these free 
parameters and is used in the evolution of the GA. This string 
is subjected to selection, inheritance, and variation. The way in which the trial 
solutions, or phenotypes, are encoded in a genotype varies from one 
implementation to the next. In our implementation, we use a decimal part 
of the real-value free parameters up to some precision (six digits by 
default). Other algorithms use the binary representation of the values 
of the free parameters, and some different encodings are possible. For 
more technical details, the history, and an example use of evolutionary 
algorithms, see \citet{eiben2015}.

Now we briefly describe the workflow of the optimizer that is based on the GA. First, the initial generation of individuals is created, 
comprising completely random solutions to the given problem. Usually,
several tens to one hundred solutions in a generation are enough. We therefore opted 
for 50 individuals. It is advisable to keep the parameter space as wide 
and open as possible to avoid losing any prospective solution. However, 
because we are only interested in physically plausible solutions, we placed
some reasonable constraints on the random solutions. In our
specific case, the random solution is a complete solution of the
fragmentation, ablation, and radiation of a fireball for the whole observation time. It contains the manually found gross fragmentation
times, fragment, and dust masses, it can include erosion of fragments, 
and we also placed constraints on the possible number of daughter fragments 
from a parent fragment and its descendants given by previous experience 
with fireball light curve and dynamics modeling \citep{borovicka2020} and 
by the requirement of keeping the number of fragments in the whole 
solution low. For each random solution, the program calculates 
its fitness.

Based on the fitness, the individuals are ordered from the best to the 
worst, and from the best individuals, pairs are created that serve as parents 
for the next generation of offspring. The parents' genomes are then cut 
in half, and the first half from the first parent is combined with the 
second half from the second parent to create a genome of one offspring. 
The other respective halves of the parent's genomes give rise to the 
second offspring. This operation is called a crossover. Details of this
operation can vary, that is, we could have a different number of parents and a
different number of crossover points, and we could combine them 
differently. Our algorithm randomly chooses between a one-point and
a two-point crossover, where the genomes are cut into three parts (of a
random length) and the central part is exchanged by the parents. This
provides the possibility to preserve some advantageous combinations of
free parameter values and therefore faster convergence.

After the crossover operation, the offspring genomes are subjected to
a mutation, that is, a random point or several points of their genome
are randomly changed. This provides the necessary quantity of variation,
which is useful when the algorithm becomes stuck in some local extreme of
the optimized function. Moreover, the mutation provides a vital source of
slower variation of free parameter values. The mutation rate (the
probability that some mutation occurs) can be adapted based on the
variance of the fitness function values among the generation. This again
serves as a corrective when the algorithm becomes stuck in a local extreme.
More technical details are given in \citet{charb2002ga}.

Using this approach, a whole new generation of individuals is created,
and the old generation is replaced, with them possibly keeping the best 
individual (a property called elitism). This process is repeated, which 
is called evolution, until some predefined criterion is reached. This can 
be a certain number of generations (we used this), a negligible change in $\chi^2$ 
value, or even some predefined $\chi^2$ value reached by the
optimization.

\subsection{GA implementation}\label{ga-implementation}
Many commercial programs implement GA or
other evolutionary strategies. However, there is also an open-source
program that has already been used in many areas of astrophysics. It is
called PIKAIA \citep{charb1995pug}. We used this well-documented and debugged piece of code with 
some minor changes that we detail below. Here we describe it briefly.

\subsubsection{PIKAIA}\label{pikaia}
PIKAIA\footnote{The source code for PIKAIA and some of the cited 
references can be found at \url{https://www.hao.ucar.edu/modeling/pikaia/pikaia.php}.} 
was developed by Paul Charbonneau and Barry Knapp in 1995. It
is written in Fortran 77 \citep{charb1995pug}. This open-source 
implementation of GA can be used for
many global optimization problems. The interface is similar to any
standard Fortran language library, such as in LAPACK \citep{anderson1999}
or the Numerical Recipes in Fortran 77 \citep{press1992}. It has been
successfully used in many research areas of astrophysics 
\citep{charb1995}. To use this program for a specific problem, the user 
has to write their own function to optimize. In 2002, the program was 
updated \citep{charb2002rel} and a thorough introduction to the use of 
GA for solving astrophysical problems was written 
\citep{charb2002ga}. The detailed documentation also contains a few 
testing problems and examples. Moreover, the program was 
transliterated to IDL\footnote{\url{https://www.l3harrisgeospatial.com/Software-Technology/IDL}} and was rewritten to more modern versions of Fortran 
(Fortran 90 and Modern Fortran).

\subsubsection{Parallel PIKAIA via MPI}\label{parallel-pikaia-via-mpi}
The most expensive part of the computation is the fitness evaluation
for every individual solution to the problem. Because in each generation,
we need to calculate 50 to 100 individual solutions, it is natural to
think about distributed computing. It enables the distribution of jobs 
between many processors that are available in large computing clusters, and
the individual solutions are computed in parallel. This approach enables 
cutting the total computation time to a fraction when compared to a
single-processor computation.

A parallel version of the PIKAIA program fortunately exists as well. It
was developed by Travis Metcalfe in 2001 \citep{metcalfe2001phdt} to 
model the internal structure of white dwarfs \citep{metcalfe2003}. 
There  are two versions of this program: one uses the parallel 
virtual machine \citep{geist1994} for communication between jobs, and 
the other version is written in the message-passing interface (MPI) standard
\footnote{\url{https://www.mpi-forum.org}}. We decided to use the more
contemporary standard, the MPI version, which is called MPIKAIA\footnote{
\url{https://whitedwarf.org/parallel}}.

The MPIKAIA program is constructed in the following way. The first
processor, which is denoted as rank 0 (or master), runs the main part of
the program. After some initial calculations, it dispatches jobs
(fitness calculation of individual sample solutions) to other ranks 
and waits for the results. When all these individual solutions 
are calculated, they are sent back to the master, which then calculates 
genetic operations on them (crossover, mutation, and calculation of a 
new generation of offspring). This is then repeated for a chosen number 
of generations or until another predefined criterion is reached.

\subsubsection{Assembling}\label{putting-it-all-together}
We rewrote the MPIKAIA into Fortran 90 standard so that it was
compatible with other parts of the FirMpik program. We also added more
communication between the jobs needed for a seamless run of the
optimization, and we had to adapt the run scripts to a specific cluster
we used for the calculations. To ensure that the program works correctly,
we used several test examples given in the documentation of the PIKAIA
program. The various GA controlling parameters were
tested to determine the optimum settings for production runs of the
program. The values of the parameters that we currently use (and only
occasionally change) are given in the next section.

\subsection{GA controlling parameters}\label{ga-control}
In the initial runs, we experimented with the number of individual
solutions in one generation (population size). We tried population
sizes of 50, 70, 96, and 144 individuals, running on 48, 48, 96, and 144
computing cores, respectively, and slightly better
solutions were produced by the smallest population of 50 individuals. The calculation was substantially faster on the corresponding number of
cores. Since then, we use a population size of 50 individuals. It is
also an optimum size with respect to the number of processors at a
single node of the computation cluster we mostly use for our
calculations. Another parameter of the MPIKAIA program is the number of 
generations that are calculated in a single run. This is currently our 
stopping criterion of the calculation, and it is between 800 and 1000.

We used a two-point crossover with a probability of 85\%. The
mutation probability is in the range of 0.05\%--25\% and is
variable based on the solution clustering. Larger clustering of
individual solutions might indicate that the algorithm became stuck in
some local minimum, and the mutation rate is increased so that some novel
solution could be found. The full generational replacement was employed
as the reproduction plan. Moreover, creep mutation and elitism were both
used in our calculations.

As stated above, the calculation is initiated by totally random
solutions to the problem. It is very useful to cover the parametric
space as widely as possible so that the GA can work
properly and find the global optimum. From this, it follows that the
values of the problem parameters such as the fragment mass, the number of
daughter fragments, or the erosion coefficient should have the widest
possible span.

\subsection{Optimization function and its logic}\label{optimization-function}
The most complicated task was writing the optimization function. This
function has to be able to construct a physically plausible trial
solution of the fireball fragmentation in the atmosphere without
unnecessary limitations. Constraints that are too strict usually prevent or hinder the natural GA function from finding a good solution. 

As mentioned above, the trial solution models the whole observation of a
fireball, including the gross and continuous fragmentations. The
fragmentation times are searched for manually and are fixed for all 
solutions. For all fragmentation times, we first chose a fragment to be 
broken apart. We did not break up eroding fragments. Then we decided how 
many daughter fragments the fragment was to produce. The reasonable 
maximum number of these fragments is constrained by the number of gross 
fragmentations and erosions observed in the radiometric curve. The 
general rule is that we sought a solution with a minimum total number of 
fragments. Our experiments show that a solution with a similar quality (with a 
similar $\chi^2$ value) can be found with more fragments, 
but we decided to favor a solution with the fewest fragments.

Next, for each daughter fragment, we chose its mass, which is constrained by the
mass available at the time of fragmentation. The minimum mass of a
fragment was fixed and given as a fraction of the parent fragment mass. It can be experimented with. After this, we randomly decided whether the
fragment eroded or not, and if it did, we randomly chose the value of
its erosion coefficient $\eta$. Its meaning is similar to the ablation 
coefficient describing how fast the meteoroid mass transforms into gas. 
In the case of $\eta$, we imagine that the fragment gradually loses 
mass in the form of debris.

The same procedure was also used to choose mass limits of dust
released in a gross fragmentation of the parent fragment in question.
Currently, we choose a single-mass value of dust grains with a similar
span of possible values as for erosion. The total mass of dust particles 
is the leftover in fragmentation of its parent fragment into a specified 
number of daughter fragments.

After we chose all the parameters, we calculated the whole fireball 
model as with the original version of the Firmodel program. Then we loaded 
fireball data and compared the model to the data. We calculated the 
$\chi^2$ value and its inverse, which we used as fitness to judge the 
quality of the calculated model. In this way, we proceeded with all the 
trial solutions, and this is the most expensive part of the calculation.

In the development of this function, we dealt with a few problems,
especially with setting a reasonable minimum mass of the fragment. We have
found a viable solution, but the optimization process occasionally
leads to a crash caused by a situation in which no parent fragment 
is left to break apart or hits the lower fragment mass limit.

\subsection{Initial mass and velocity}\label{init-mass-vel}
To successfully model the fireball, we first need to estimate the
initial mass and velocity of the meteoroid. The initial velocity is
calculated by fitting the linear part of the dynamics curve by a
straight line (length versus time; the velocity is the slope). This value is
usually precise enough, but we still allowed for some small changes
in the modeling process. We let the algorithm randomly choose the
initial velocity from the interval around the initial value and form the
sample solution with this value. The size of the interval at the
beginning of the process was $0.2\,\rm{km\cdot s^{-1}}$ and was decreased 
to $0.05\rm{-}0.1\,\rm{km\cdot s^{-1}}$ as the best solution is approached. 
The initial height is optimized during modeling as well to obtain a 
consistent solution of the meteoroid dynamics.

\begin{figure}
   \centering
   \includegraphics[width=\hsize]{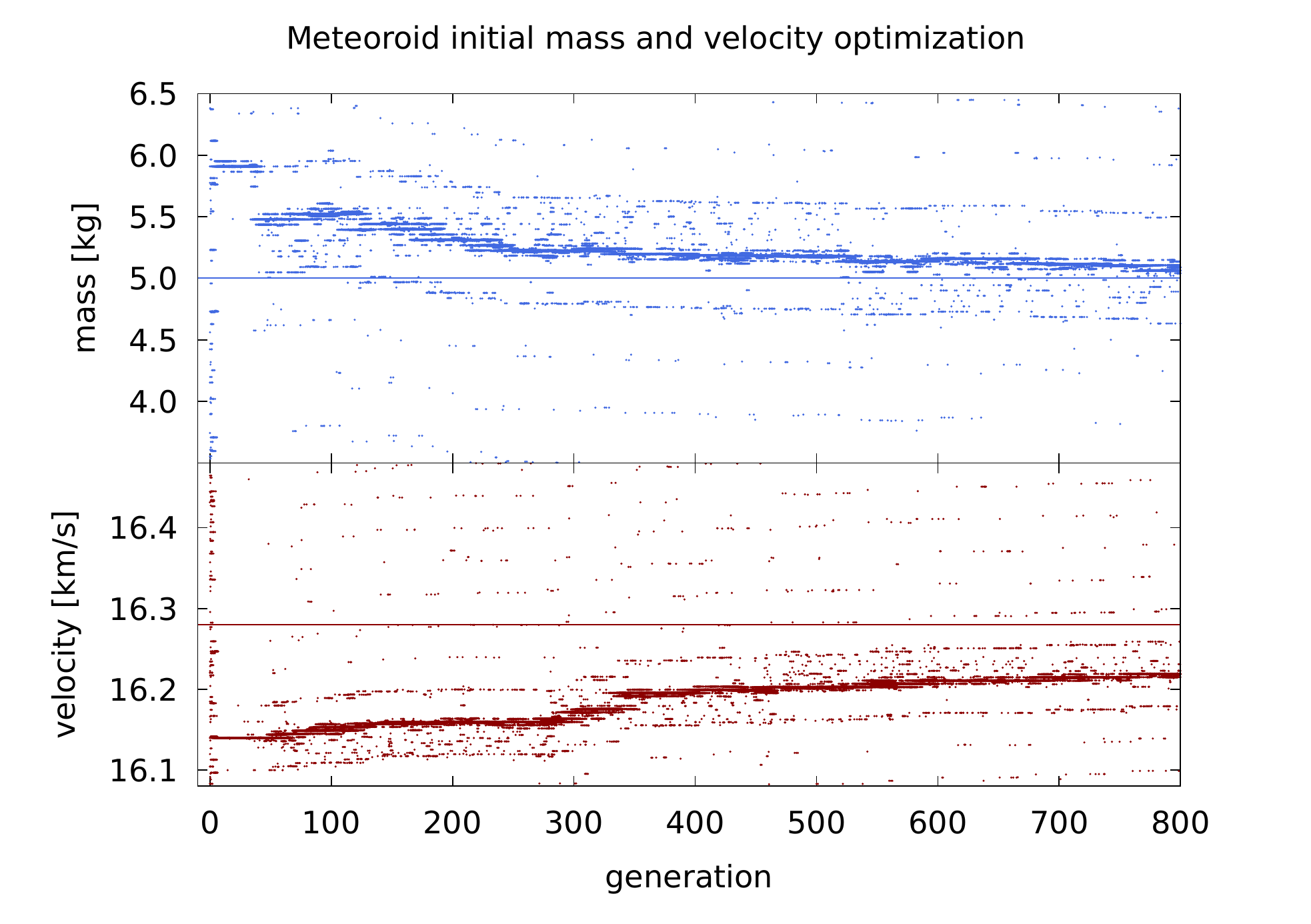}
      \caption{Example of the initial mass and velocity 
evolution of the meteoroid for the best solution of the EN020615. The solid line shows 
the starting values that were found by the procedure detailed in the text: an 
initial mass of $5.0\,\rm{kg}$, and an initial velocity of 
$16.28\,\rm{km\cdot s^{-1}}$. The width of the searched interval of 
values was set to $\pm1.5\,\rm{kg}$ for the mass and 
$\pm0.2\,\rm{km\cdot s^{-1}}$ for the velocity.}
         \label{mass_vini_opt}
\end{figure}

The initial guess of the mass was first calculated by integrating the
radiometric light curve (or a regular light curve), which yielded the 
radiated energy. Then we set the radiated energy equal to the kinetic 
energy of the body, multiplied by the assumed luminous efficiency 
($\tau$). We used the function of \citet{borovicka2020}, which depends on 
the velocity (already calculated) and the mass, which we were about to 
calculate. Therefore, we calculated the mass iteratively, that is, we 
calculated the mass with an initial guess of $\tau$, and we updated the mass and
$\tau$ until the mass converged to some value that no longer changed (usually only a few iterations). By this approach, we usually 
obtained a lower bound of the mass because the smaller fragments have 
lower $\tau$ and $\tau$ also decreases with decreasing velocity. A more 
precise value has to be calculated in the optimization process in the 
same way as the velocity. The size of the interval was initially set to 
0.5 of the mass value and was decreased as appropriate. An example of the 
initial velocity and mass convergence in the optimization process is 
given in Fig.~\ref{mass_vini_opt}.

\subsection{Search for fragmentations}\label{search-for-frags}
The current version of the FirMpik program is not able to find
fragmentation times in the radiometric curve data. Therefore, we have to
search for them manually. We marked the approximate fragmentation times, and 
then we increased its precision through an automatic process.

The gross fragmentation is usually visible in the curve as a fast or 
immediate and obvious brightening, followed by a more gradual dimming of 
the fireball. They are modeled as a sudden breakup of a parent fragment 
with a dust release. On the other hand, continuous fragmentations are 
more subtle, and some experience is required to successfully find it. We 
modeled continuous fragmentations as an erosion of dust particles from 
a single fragment. More than one fragment occasionally starts to erode at a 
given fragmentation time, with a different erosion coefficient and size
distribution of the released particles. Through this, we can fit some more
complicated shapes of the radiometric data (e.g., different slopes).

\subsection{Smoothing of the radiometric
curve}\label{smoothing-of-the-radiometric-curve}
\begin{figure}
   \centering
   \includegraphics[width=\hsize]{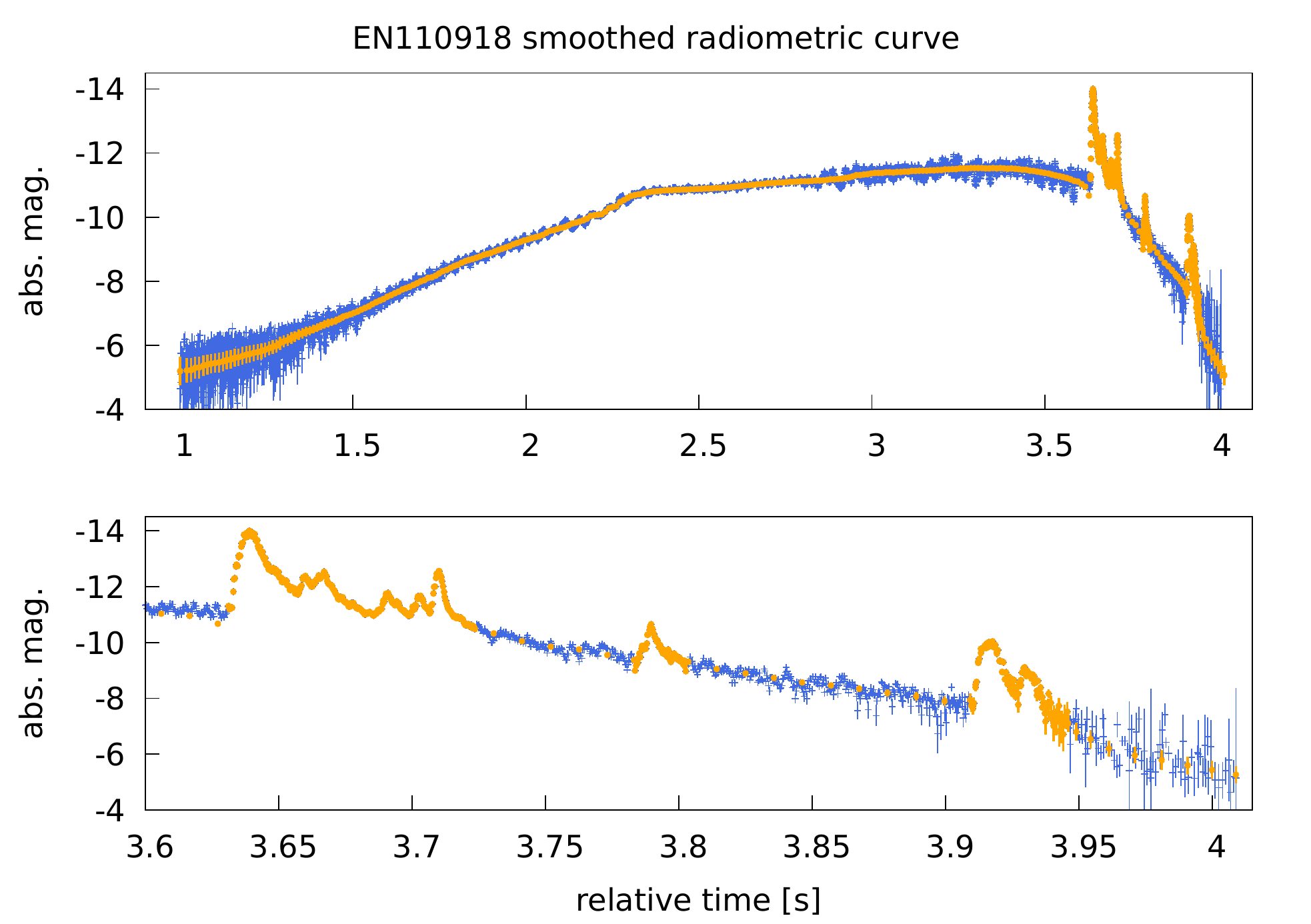}
      \caption{Smoothed radiometric curve of the EN110918 shown by 
orange dots with error impulses, compared with the original, labeled by 
blue pluses with error impulses. The upper panel shows the whole 
radiometric curve, and the lower panel shows the end of it in greater 
detail. The flares are copied from the original curve, and the other 
parts are filtered using the median running box.}
         \label{110918_smooth}
\end{figure}

Radiometric curves contain very fine details of the fireball brightness
because their sampling rate is $5000\,\rm{Hz}$. They contain semi-harmonic
brightness changes whose origin is still debated (e.g., \citep{beech2000, 
baba2004,spurny2008,spurny2012}). Our current model, however, does not 
contain a description of these rapid changes, and therefore, we needed to 
remove the changes from the radiometric curve for a successful modeling. On the 
other hand, we need to preserve all the details of the gross 
fragmentations and general trends that we model with the FirMpik 
program. We therefore manually separated the radiometric curve into 
two parts, one part that contained the data points of the flares and their close 
vicinity, and the other part contained the outside parts. The latter 
curve was then automatically smoothed by the running-box median filter 
with an appropriate box width. This procedure reliably removes 
the rapid brightness variations, but preserves the general trends in 
fireball brightness. The parts of the curve that contain the gross 
fragmentations remain unchanged. An example of a smoothed radiometric 
curve is shown in Fig.~\ref{110918_smooth}.

Moreover, for a successful automatic modeling, it is not practical to work
with such a high sampling rate of the radiometric curve, unless we wish
to model its very fine structure. Therefore, we decided to automatically 
dilute the part of the radiometric curve that lies outside the gross 
fragmentations. Another reason is that in modeling, we wish to have a 
comparable number of data points for each data set, that is, for the 
radiometric curve and the light curve, and for the dynamics as well. It helps 
the algorithm to fit the data with a similar priority.

For each fireball, we needed to test the appropriate box width for
the running-box median filter. Therefore, we calculated the smoothed
curve with different box widths and then selected the width that captured all
the main features of the original radiometric curve.

\subsection{Dataset weights}\label{dataset-weights}
The first attempts to automatically model the fireball data were not
very successful. For instance, we obtained a perfect fit for the dynamics of
the fireball and a much poorer fit for the radiometric curve and 
light curve. This was probably caused by an unknown conversion between 
brightness and distance contributions to the $\chi^2$ sum (the inverse of 
which is our fitness), or simply put, by a magnitude--kilometer conversion 
factor. Moreover, our datasets usually have a different number of data 
points, and those data have also different uncertainties. Therefore, we
experimentally set different weights to each dataset in the $\chi^2$ value
calculation to show the algorithm what to focus on when
individual solutions were evaluated. After some experimenting with the specific weight
values, we were able to obtain a more reasonable fit for both datasets.
The specific setting of the weight is still rather empirical, and it is a
source of uncertainty when we start calculating the new fireball. The
current approach is to run several individual calculations with a wide
range of weights, to choose those that are more viable, and to gradually
narrow down the span of the weights. The current construction of the
optimization process does not allow setting the weights as a free
parameter.

\subsection{Largest impact on the $\chi^2$ sum}\label{chi2-impact}
To keep the number of free parameters low, we tested which of 
them were the most important and which were less important. 
We fixed all the parameters except for one, and 
inspected the change in the $\chi^2$ sum. Thus we found that the number of fragments had largest 
impact on the $\chi^2$ value, especially 
the fragments that were released from the meteoroid in the first fragmentation. The second 
most important parameter was an erosion of some of the released 
fragments. The erosion controls the overall shape and slope of the total 
brightness curve in the model. The third most important group of 
parameters were the fragment masses and the dust mass. The dust mass distribution limits were less important because they mainly affect 
the shape of the flares, but the flares are usually short and therefore 
have little effect on the total $\chi^2$ sum.

\subsection{RLC weights}\label{rlc-weights}
To further emphasize the importance of gross and continuous
fragmentations within the radiometric curve, we decided to set higher
weights for the vicinity of the fragmentation times in the $\chi^2$ sum
calculation. Using this approach, we can tell the algorithm which parts 
on the radiometric curve have to be fit more precisely and where the 
fit can be looser.

\subsection{Modeling outputs}\label{modeling-outputs}
The fireball modeling yields several important parameters describing the
fragmentation of the meteoroid in the atmosphere. These are the number
of fragments in each fragmentation time, their respective masses,
whether or not the fragment erodes, its erosion coefficient, and the
mass limits of the eroded dust grains. Moreover, we determine the amount of
dust released in gross fragmentations and schematically determine the mass
distribution of the dust. The model optimizes the initial
velocity and mass of the meteoroid, so that we obtain more precise values than we can calculate 
from the dynamics and radiometric curve alone. The program can also 
resolve whether some meteorites may be found on the ground and 
 their number and respective masses.

From the model, we deduced the height of the fragmentation for
each fragmentation time. Based on this, we can calculate the atmospheric
density at that height. Using the measured velocity at the same time, we
calculated the dynamic pressure ($p=\rho v^2$) acting on the front
part of the meteoroid. This is a proxy for a meteoroid shear strength 
\citep{robertson2017}, and it is a very important mechanical parameter of 
the meteoroid material. In this way, we can compare different meteoroid 
materials and in turn obtain other constraints on the properties of its parent body. More clues about the origin of the meteoroid can be deduced from 
the whole fragmentation process (how it breaks up, whether it erodes or 
not, etc.). The data on fragment masses and their strength approximation 
(just before their fragmentation) can then be plotted in log--log plane, 
as shown for the test fireballs presented in this paper in 
Sect.~\ref{results}. In this way, we can also compare different solutions of 
a single meteoroid fragmentation.

Because of the extensive use of pseudo-random numbers, the algorithm is
inherently stochastic. To obtain a reliable fit of the fireball data, we
need to run the calculation several times to some tens of times and
then process the results statistically. The number of runs depends on
the number of data at hand, on their complexity (e.g., the number of 
fragmentation times and other features of radiometric data), and thus mainly on the number of free parameters we need to find. The number 
of free parameters is proportional to the number of fragments that are
produced during meteoroid fragmentation,

\begin{equation}
N_{\rm parm} = 5\cdot N_{\rm frag} + N_{\rm dust} + 2
.\end{equation}

For each run, we typically obtained several hundred best-fitting solutions 
with almost the same $\chi^2$ value that are nearly identical. In two
different runs, the fits were of comparable quality, but the details
of the fragmentation of the meteoroid can differ. From the set of the 
best solutions, we can estimate the uncertainties, that is, we can calculate the 
most probable number of fragments in each fragmentation time, 
the masses of individual fragments, and so on.

\subsection{Uncertainties}\label{uncertainties}
The initial idea of calculating the uncertainties of (some of) the free model parameters was to calculate the $\chi^2$ value threshold, which is an equivalent of a one sigma confidence interval for a single measured value by a standard statistical theory. However, this was not possible in our case mainly because the modeling problem is 
not linear in those parameters \citep{bevington2003,andrae2010}. As 
noted above, the 
reduced $\chi^2$ value therefore does not have the usual statistical 
meaning and cannot be used to calculate the probability of finding the 
optimized parameter in a certain interval around the value found by the 
optimization.

To find some equivalent of the uncertainties of the quantities obtained 
in the simulations, we decided to use the approach of \citet{gibson1998}. 
The uncertainties of the initial velocity and mass of the meteoroid were estimated 
based on the set of (some hundred) acceptable solutions with a 
$\chi^2$ value lower than some chosen threshold value. In our case, it 
was arbitrarily set to 1.1 of the $\chi^2$ value of the respective 
acceptable solution (each acceptable solution had a different minimum 
$\chi^2$ value). In this way, we constructed histograms of initial 
masses 
and velocities that were indicative of a confidence interval. It would 
probably require many more successfully converged simulations for a 
single fireball to numerically describe the underlying (complicated) 
distribution of the optimized quantities and to derive a statistically 
meaningful confidence interval for those quantities as an interval 
between, for instance, the 5th and 95th percentile. However, this is too 
 expensive computationally.

Another possibility is to calculate the uncertainties using the Monte Carlo 
method \citep{press1992}. This requires using the current datasets with 
their intrinsic uncertainties, to prepare many hundreds to thousands of 
synthetic datasets within these data uncertainty intervals, and then 
optimizing these datasets only to obtain a sampling of the free model 
parameters, which enables us to derive their uncertainties. This is 
perfectly viable when data with a moderately complicated model need to be fit 
that takes seconds to minutes to calculate. An order-of-magnitude 
estimate of the computation time for our problem, however, gives some 
thousands to tens of thousands of hours of computation time, which is 
an unreasonably long time. Therefore, we decided to present 
the uncertainties of some of the free model parameters only in the form 
of plots showing the best solutions without a more rigorous 
quantification (see Sects.~\ref{amass_vini_unc} 
and~\ref{pressure-mass-plot}).

\section{Results}\label{results}

\subsection{Manual modeling comparison}\label{manual-comp}
The correct function of the algorithm was tested by a blind
test that was constructed in the following way. One of us (JB) chose five previously modeled fireballs with abundant data taken with
the EN that were described by \citet{borovicka2020}. The other author
(TH) then prepared the data for the automatic modeling and ran a
different number of simulations for each fireball. The number of 
simulations spanned from 10 to 70. One reason for the high number 
of simulations for each fireball is the stochasticity of the algorithm, 
which was initialized randomly, and all the genetic operations are also 
probabilistic and were calculated with pseudo-random numbers.

The simulations were evaluated, and the best solutions were chosen
based on the reduced $\chi^2$ value and the overall look of the fits
of the radiometric curve, photometric light curve, and dynamic data. The 
best solutions were then compared to the manual solutions of 
\citet{borovicka2020}. In the following, we describe the solutions for 
the five fireballs in detail. The figures for four fireballs are presented 
in Appendix~\ref{app_fig}. The name of a fireball observed by 
the EN contains the EN prefix, followed by the date and time (UT) of the 
fireball observation start in the format ENddmmyy$\_$hhmmss. We give the 
full name for each fireball first, but then we use a short form of 
ENddmmyy to refer to it. 

\paragraph{EN260815$\_$233145}\label{en260815}
This fireball was caused by a $6.5\,\rm{kg}$ meteoroid traveling at 
$19.3\,\rm{km\cdot s^{-1}}$. First, a very gradual brightening of the 
fireball was observed, while at the same time, there were very fast 
semi-periodic variations in the radiometric curve. In the second 
half of the curve, three very bright flares were observed (the first 
flare was the brightest at $-11.7\,\rm{mag}$), and one not very 
conspicuous flare occurred at the end. These flares served as unequivocal 
fragmentation times in the modeling. 

This fireball was the first that we attempted to model with the 
FirMpik program. It also served as a model case for developing the 
optimization function and for thorough testing of the code that 
preceded the production runs of the following fireballs. Because the 
fragmentation points are more or less obvious in the radiometric curve, 
the modeling was relatively easy. The hardest part in modeling this and also other fireballs is to reconcile the radiometric curve with 
the dynamic data at the very end of observations. This is one of the 
reasons we had to calculate several (tens of) runs.

\begin{figure}
   \centering
   \includegraphics[width=\hsize]{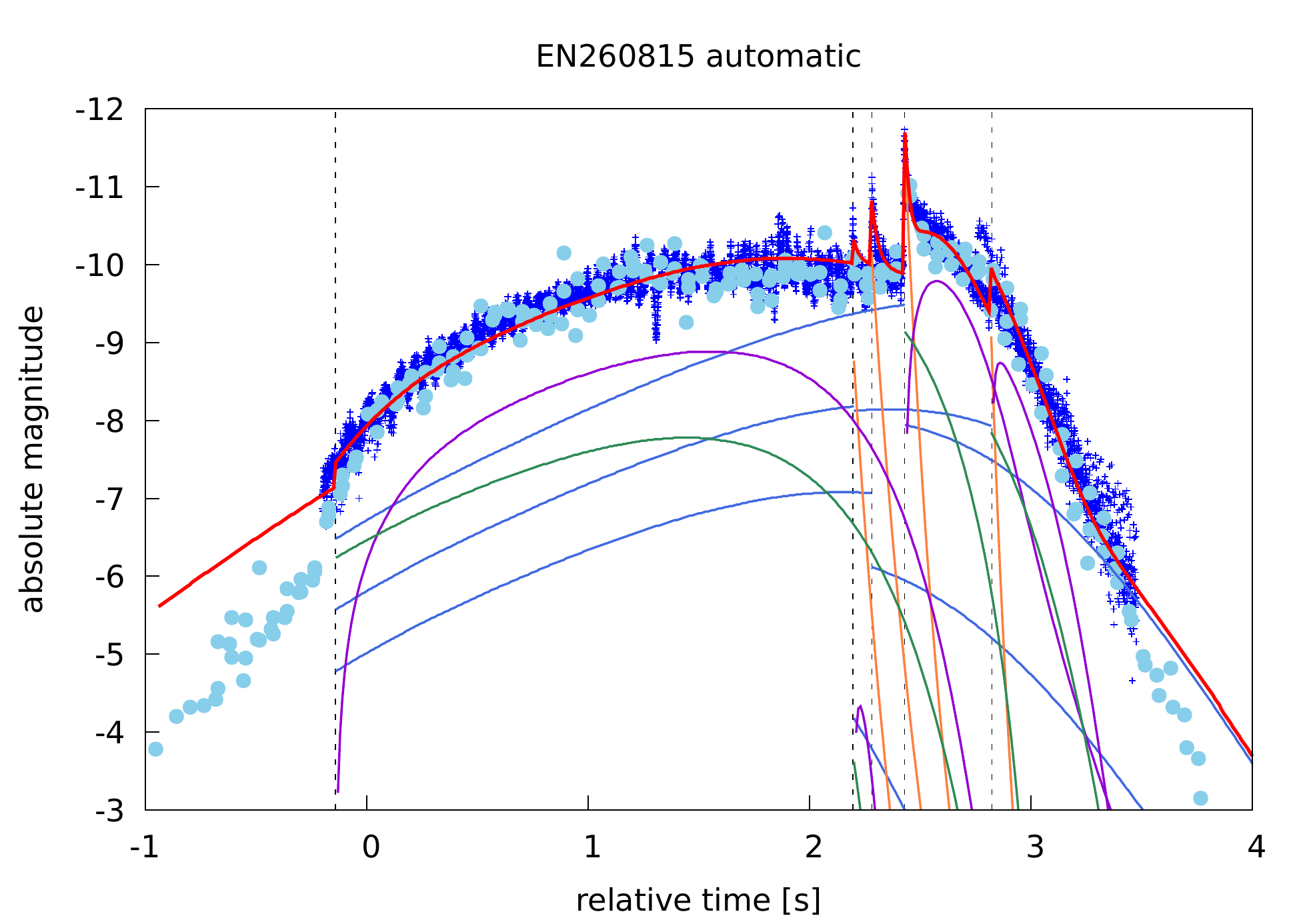}
      \caption{Automatic solution of the EN260815 fragmentation 
compared to the observed radiometric curve (dark blue pluses) and a 
photometric light curve (sky blue disks). The total model brightness is 
shown as a solid red line, the brightness of regular fragments is 
shown as blue curves, green curves signify eroding fragments, violet 
lines indicate dust particles released from these fragments, and orange 
curves denote regular dust released in gross fragmentations. 
Fragmentation times are shown with vertical dashed lines.}
         \label{260815_best_rlc}
\end{figure}

\begin{figure}
   \centering
   \includegraphics[width=\hsize]{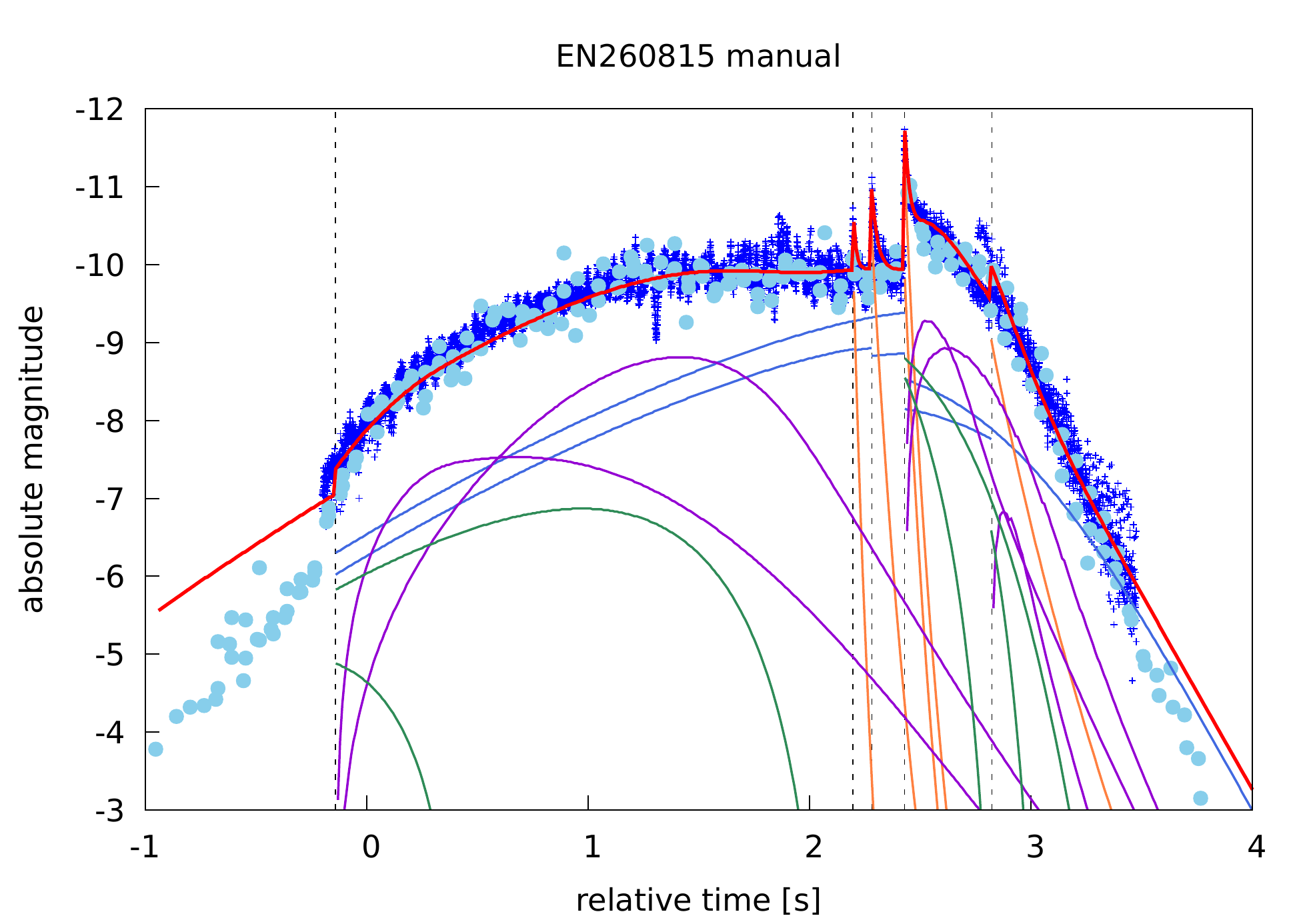}
      \caption{Manual solution of EN260815. Labels are the same as in 
Fig.~\ref{260815_best_rlc}.}
         \label{260815_man_rlc}
\end{figure}

\begin{figure}
   \centering
   \includegraphics[width=\hsize]{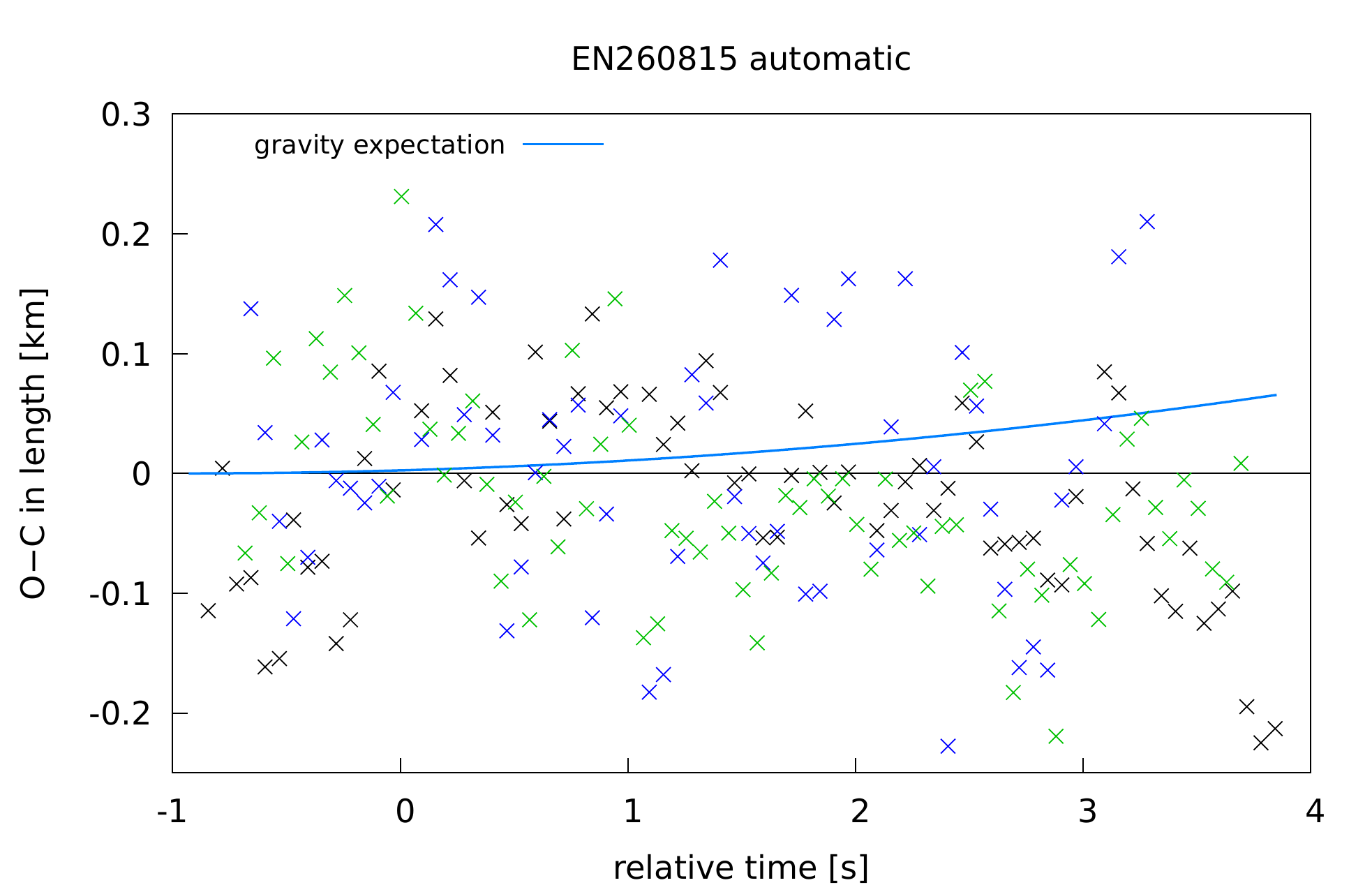}
      \caption{Residuals of length for an automatic solution of the 
EN260815. Different colors indicate the stations of the EN that were used to calculate the model. The solid blue line predicts the acceleration of the 
meteoroid by gravity.}
         \label{260815_best_len}
\end{figure}

\begin{figure}
   \centering
   \includegraphics[width=\hsize]{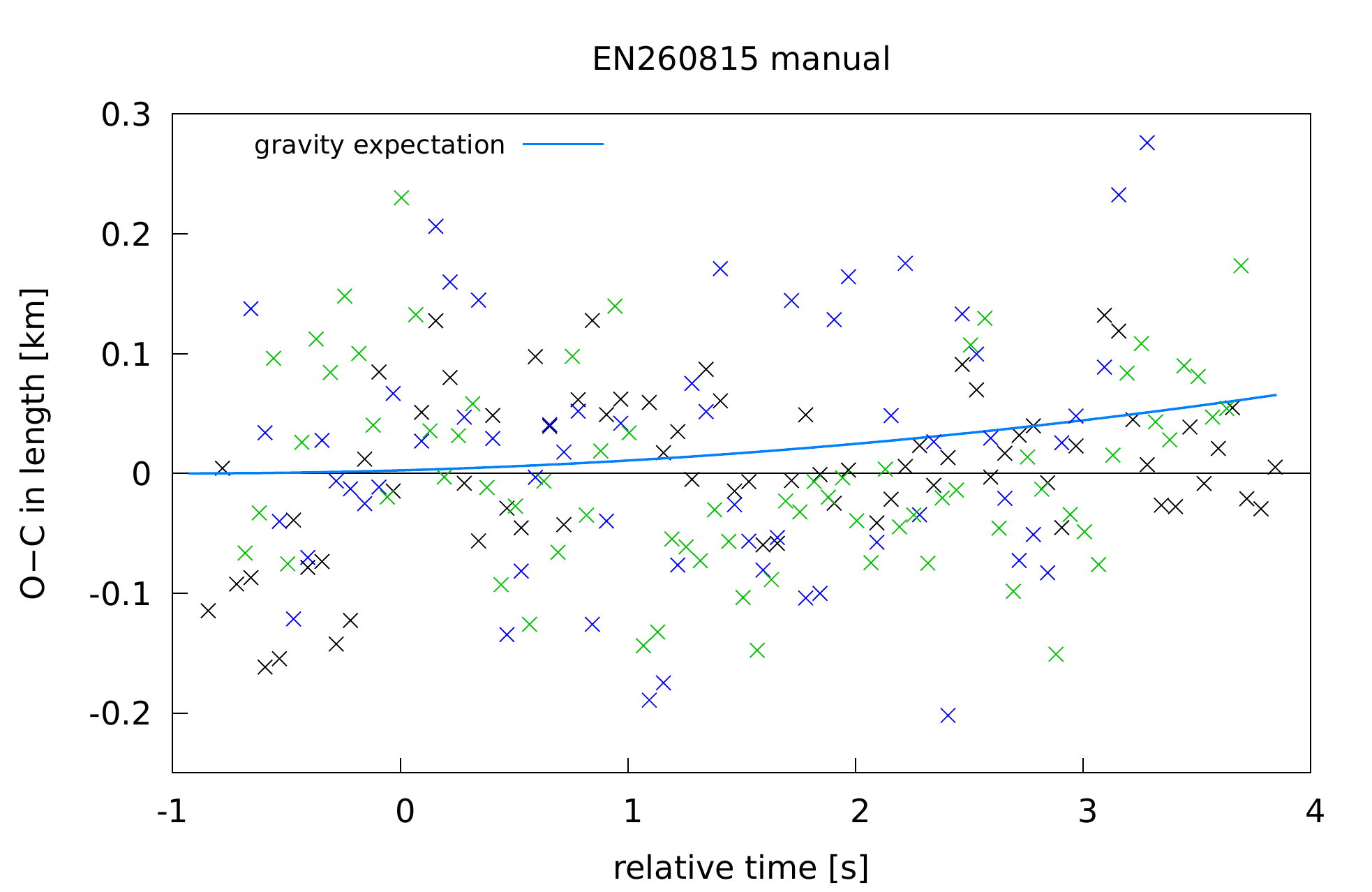}
      \caption{Residuals of the length for a manual solution for 
EN260815. Labels are the same as in Fig.~\ref{260815_best_len}.}
         \label{260815_man_len}
\end{figure}

The automatic solution in Fig.~\ref{260815_best_rlc} is very similar to 
the manual solution in Fig.~\ref{260815_man_rlc}, although the flares are not 
fit with the same precision and the model is slightly brighter at the 
end of observations. Residuals of dynamics fit are shown in 
Figs.~\ref{260815_best_len} and \ref{260815_man_len} for the automatic and 
manual solutions, respectively.

As we mentioned above, the dynamic data only refer to the foremost 
fragment, but it is a very important piece of information. While 
the radiometric curve alone can be fit by a large number of fragment 
combinations, the dynamic data constrain the plausible combinations because they 
sensibly check the size of the foremost fragment (smaller fragments 
decelerate more than larger ones). It is also an independent check on 
whether we correctly found the fragmentation times.

We expect the points in the O$-$C plot to be randomly and 
symmetrically distributed around the blue gravity expectation curve 
(because gravity accelerates the meteoroid and its fragments). There should be no trends or periodic behavior, and a reasonable 
dispersion of the points is on the order of a hundred meters for 
high-quality data.

\begin{figure*}
   \centering
   \includegraphics[width=\hsize]{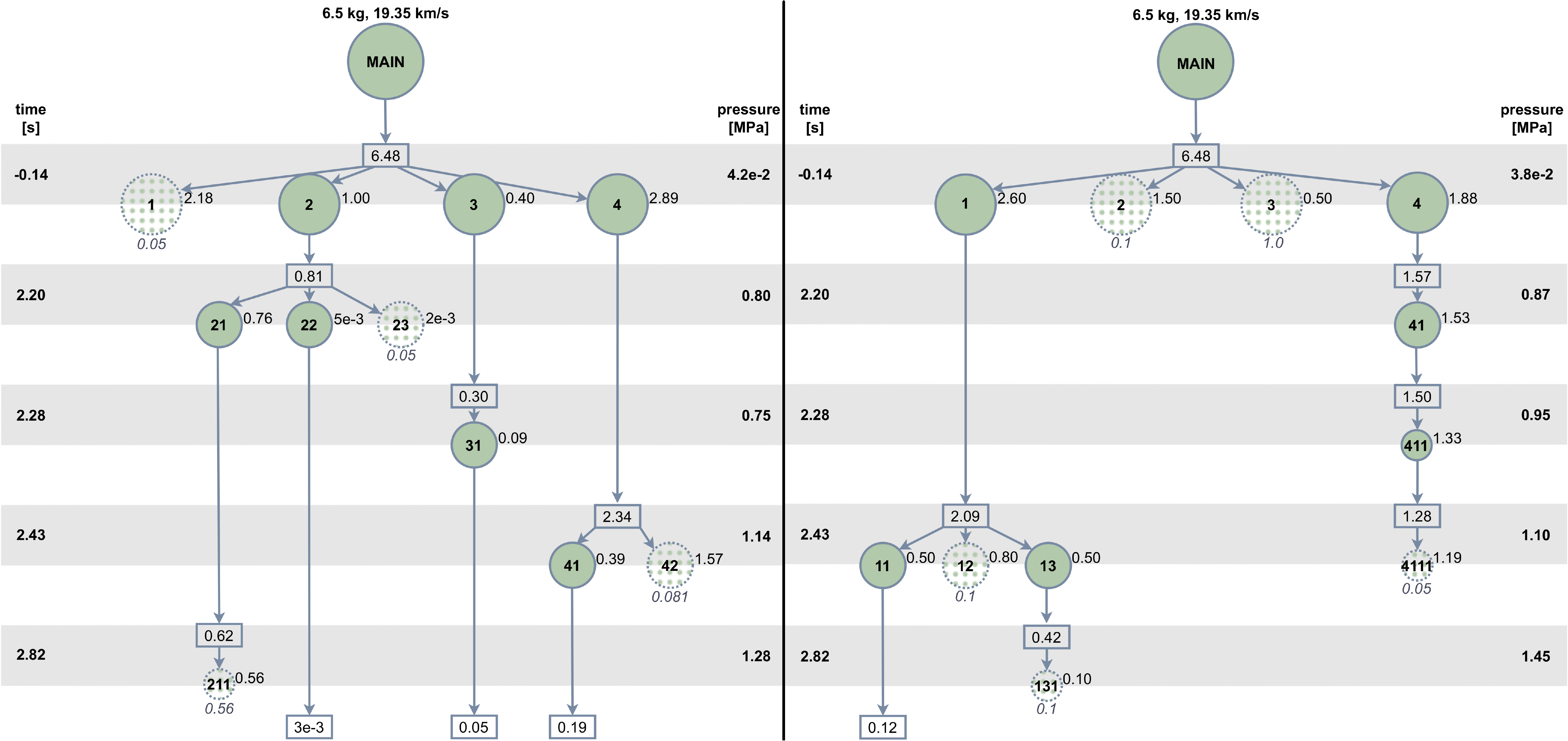}
      \caption{Comparison of the automatic (left panel) and
manual (right panel) solution for EN260815. The solid disks are 
regular ablating fragments, and the spotted disks with a dotted border 
are eroding fragments. The time runs from top to bottom, numbers in the 
disk center are fragment names, the number to the right of the disk is 
the initial mass of the fragment in kg, and the number at the bottom of the disk is 
the fragment erosion coefficient $\eta$ in $\rm{kg\cdot MJ^{-1}}$. The 
final mass (also in kg) of the fragment before its fragmentation is shown as 
blue rectangles. Slight differences in dynamic pressures are caused by 
two different models of atmosphere in these two solutions.}
         \label{260815_comp}
\end{figure*}

The two solutions are also compared graphically in 
Fig.~\ref{260815_comp}, where the disruption cascade is depicted schematically. The cascades differ, but the eroding versus ablating mass 
ratio is similar at any instant displayed in the plot in both 
solutions. The first model predicts a mass that is twice higher in meteorites 
than the second model (0.24\,\rm{kg} vs. 0.12\,\rm{kg}), which corresponds to 
the higher modeled brightness at the end.

\paragraph{EN110918$\_$214648}\label{en110918}
This relatively bright fireball was caused by a $6.5\,\rm{kg}$ meteoroid 
traveling at $23.5\,\rm{km\cdot s^{-1}}$. The radiometric curve begins 
with a gradual brightening that takes almost $1.25\,\rm{s}$ and is 
followed by a plateau. At $t = 2.8\,\rm{s}$, some brighter 
semiperiodic brightness variations appear in the curve. These are 
followed by a series of nine flares that signify important breakups of 
the meteoroid. Four of them are very bright (the brightest reached 
$-14.25\,\rm{mag}$). In the end, the overall brightness of the fireball 
decreases very rapidly.

This fireball was among the most difficult to model. It was 
straightforward to find the fragmentation times because they are indicated 
by very bright and short flares in the radiometric curve. In the first 
step, we used a moderate time resolution of $0.01\,\rm{s}$, and after 
several attempts, we successfully modeled the fireball 
(Fig.~\ref{110918_low_timeres_rlc}). At this time resolution, however, 
some of the bright flares (gross fragmentations) were heavily smeared. 
Therefore, we increased the time resolution to $0.005\,\rm{s}$, and after 
varying the weights and also fixing the initial velocity and 
initial mass of the meteoroid found in the previous modeling, we were 
able to find a good solution with this higher time resolution as well 
(Fig.~\ref{110918_best_rlc}). The manual solution fits the radiometric 
curve even better with more details (Fig.~\ref{110918_man_rlc}). The length 
residuals shown in Figs.~\ref{110918_best_len} and \ref{110918_man_len} 
are almost the same. In this case, we did not use photometric data to construct the 
model, and therefore they are not plotted in the figures.

\paragraph{EN240217$\_$190640}\label{en240217}
This fireball was caused by a meteoroid of $2.5\,\rm{kg}$ traveling at 
$17.8\,\rm{km\cdot s^{-1}}$ with extremely high semiperiodic 
variations in brightness in the radiometric curve. On the other hand, 
we did not observe any obvious gross fragmentations that usually 
manifest themselves in the curve by a sudden brightening on the order 
of some tenths to several magnitudes. Therefore, the fragmentation times 
used in the modeling are less certain, and they were needed to sensibly 
fit the overall shape of the radiometric and photometric data as well 
as the dynamics of the meteoroid. The automatic solution 
(Fig.~\ref{240217_best_rlc}) is very good and comparable to the manual 
solution (Fig.~\ref{240217_man_rlc}), except for one detail: In the 
manual solution, \citet{borovicka2020} attempted to model some of the 
brightenings at the beginning as gross fragmentations. However, the 
overall mass loss in these fragmentations was negligible, and they had 
little effect on the latter parts of the light curve and on the 
dynamics. The residuals of the dynamic fit were somewhat better for the 
automatic solution, especially in the time interval $3\rm{-}4\,\rm{s}$ 
(Figs.~\ref{240217_best_len}~and~\ref{240217_man_len}).

\paragraph{EN020615$\_$215119}\label{en020615}
This fireball was caused by a $5\,\rm{kg}$ meteoroid traveling at 
$16.2\,\rm{km\cdot s^{-1}}$. The brightness variations of the fireball 
are present in the radiometric curve from the very beginning. The 
initial increase in brightness is very fast and progressive, and at time 
$2\,\rm{s}$, the fireball reaches maximum brightness, experiencing 
large semiperiodic variations in brightness at the same time. A brightness plateau follows, and after $t = 3\,\rm{s}$, the 
brightness starts to drop gradually. As usual, there is a series of 
flares in the radiometric curve: We observe five of them, while 
the overall brightness of the fireball decreases gradually. The 
automatic (Fig.~\ref{020615_best_rlc}) and manual 
(Fig.~\ref{020615_man_rlc}) solutions are similar, except for the 
beginning, where the first fragmentation time was chosen differently and 
the manual solution used only one fragmentation. This resulted in a poorer 
fit. The length residuals have a similar dispersion in both cases 
(Figs.~\ref{020615_best_len}~and~\ref{020615_man_len}), but their 
distribution at the beginning in the manual solution is more even.

\paragraph{EN180118$\_$182623}\label{en180118}
The initial mass of the meteoroid causing this fireball was slightly lower 
than $3\,\rm{kg}$, and it traveled at $20.0\,\rm{km\cdot s^{-1}}$. 
The first half of the radiometric curve is very smooth, without any 
significant semiperiodic variations in brightness, while the second 
half of the curve is quite the opposite. At the time $3.0\,\rm{s,}$ 
there is a drop in brightness that cannot be explained by the current 
model, and we therefore decided to ignore it in the modeling process. The model 
only roughly fits the brightness in that place. Then the fast 
variations in brightness appear in the curve, and they completely 
disappear at time $3.85\,\rm{s}$, reappearing again at $4.1\,\rm{s}$. 
No flares are apparent in the radiometric curve, which 
complicates the modeling. We decided to estimate the fragmentation 
times based on changes in brightness trends in the radiometric curve. 
This led to a less satisfactory model fit than in the previous cases. The end of the radiometric curve in particular is poorly fit 
(Fig.~\ref{180118_best_rlc}). The manual solution is better 
(Fig.~\ref{180118_man_rlc}), mainly because of the better fit of the 
aforementioned brightness drop and also because of the more extensive use of gross 
fragmentations between $\sim3.8\rm{-}4.6\,\rm{s}$, which allowed the 
substantial mass loss of the meteoroid and in turn a better fit. The 
automatic algorithm is not able to use this process when it is not 
clearly present in the data, because it optimizes the $\chi^2$ sum with 
respect to the median radiometric curve, and these departures would 
cause a higher $\chi^2$ sum and therefore lower fitness value. In the 
future, we wish to address this problem with a more versatile fitness 
value or a different $\chi^2$ calculation. The length residuals are 
then shown in Figs.~\ref{180118_best_len} and \ref{180118_man_len} .
They support the quality comparison of the two solutions. Both of them 
have a less satisfactory fit around the time $4\,\rm{s}$.

\subsection{Initial mass and velocity uncertainties}\label{amass_vini_unc}
\begin{figure}
   \centering
   \includegraphics[width=\hsize]{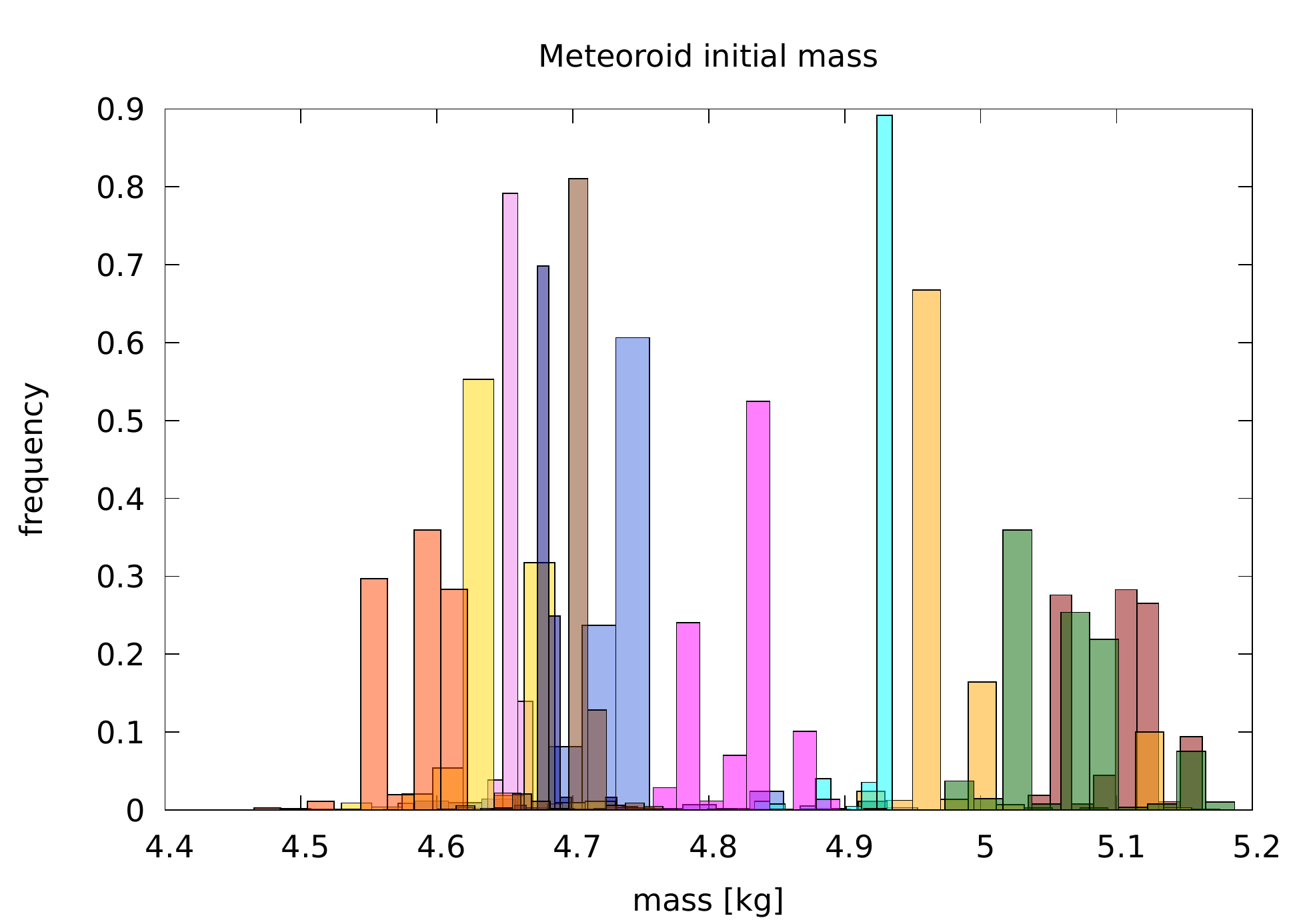}
      \caption{Spread of  the initial mass for EN020615. The spread 
displays its uncertainty. The plot shows the 11 best solutions and 
their vicinity with bars of different colors.}
\label{hist_amass}
\end{figure}

\begin{figure}
   \centering
   \includegraphics[width=\hsize]{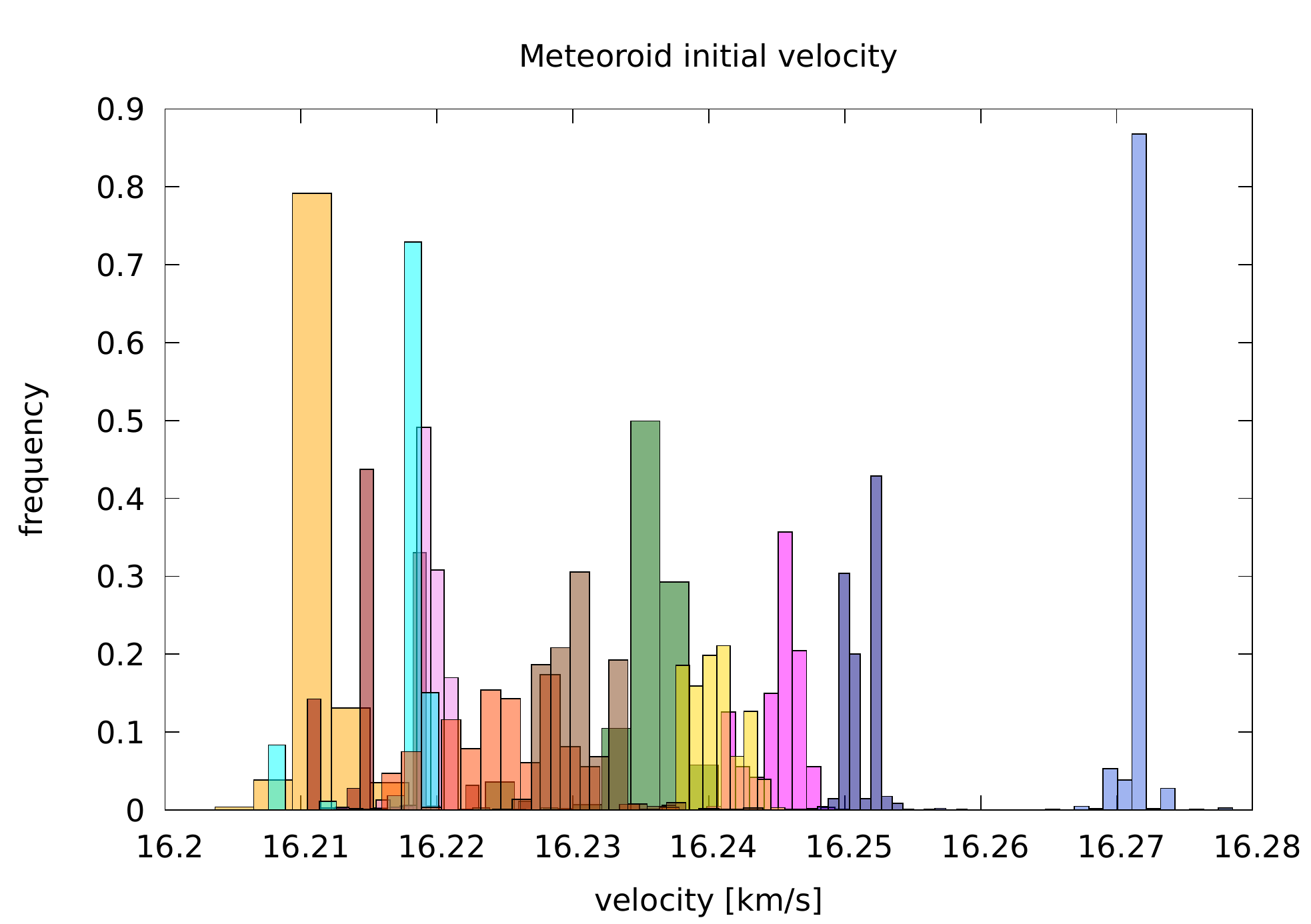}
      \caption{Spread of the initial velocity for 
EN020615. The spread displays its uncertainty. The same color as in 
Fig.~\ref{hist_amass} marks the same solution.}
\label{hist_vini}
\end{figure}

The uncertainties of the initial mass and velocity of the meteoroid are presented 
in Figs.~\ref{hist_amass} and~\ref{hist_vini}. They show the 
spread of the respective values for the 11 best solutions of the 
EN020615 fireball found by the FirMpik program. Each histogram shows with a 
specific color the distribution of  initial mass 
(velocity) of the meteoroid for a single solution and its close vicinity as measured by 
the $\chi^2$ value. The width of the histogram columns reflects the 
spread of the mass (velocity) values for each best solution (the number 
of histogram columns was fixed). The 
shape of individual histograms is caused by the nature of the GA 
approach to finding a solution. The spread of the initial mass and 
velocity values and the overall shape of all histograms in 
Figs.~\ref{hist_amass} and~\ref{hist_vini} suggests a complicated 
distribution of the uncertainties. The details of this procedure were described 
above in Sect.~\ref{uncertainties}. For other fireballs, the 
uncertainty distribution looks similar. For comparison, the manual 
solution of the EN020615 gave an initial mass of the meteoroid of 
$5.0\,\rm{kg}$ and an initial velocity of $16.28\,\rm{km\cdot s^{-1}}$.

\subsection{Pressure--mass plot}\label{pressure-mass-plot}
\begin{figure}
   \centering
   \includegraphics[width=\hsize]{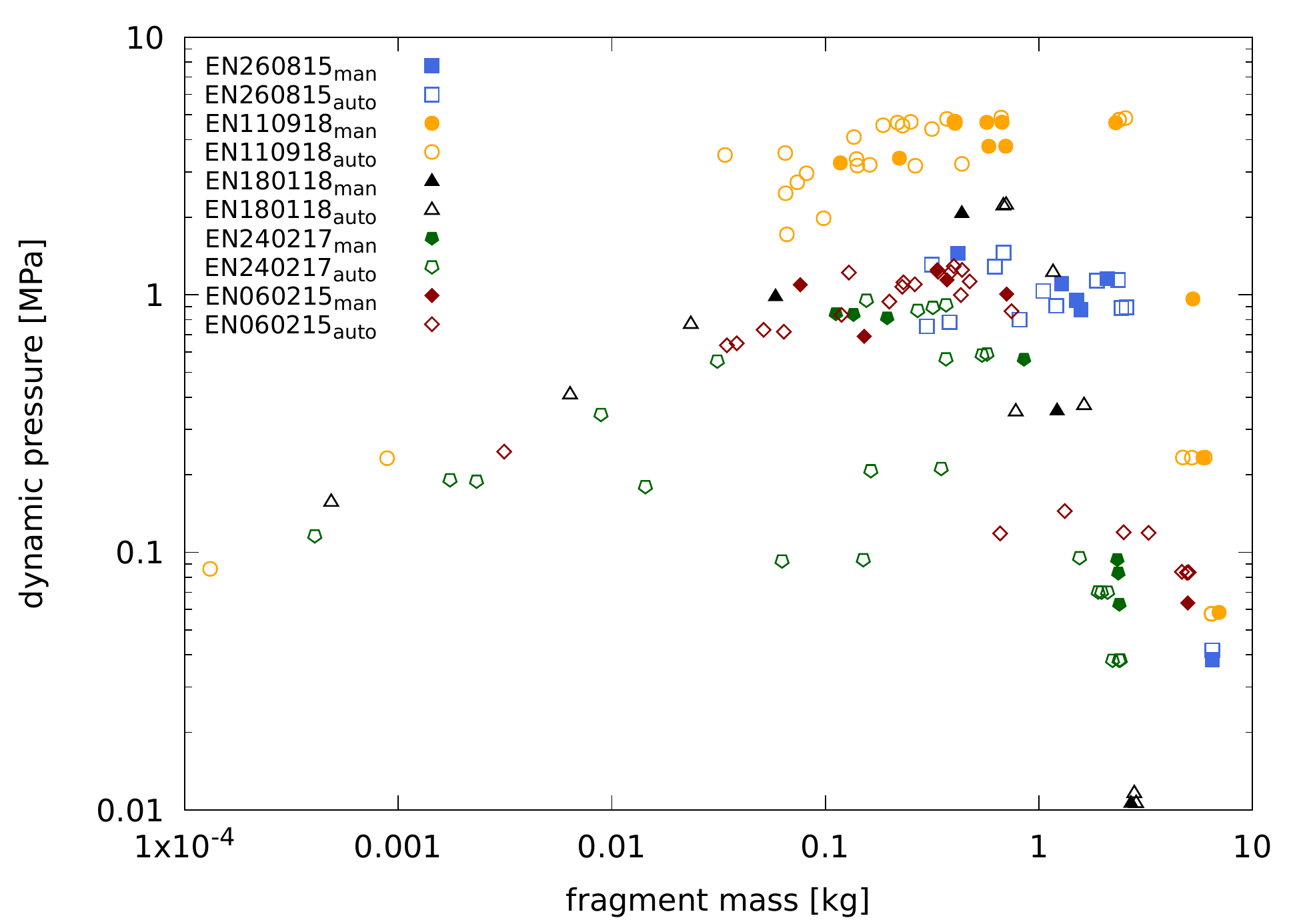}
      \caption{Comparison of automatic (open symbols) and manual 
(solid symbols) meteoroid fragmentation modeling in dynamic 
pressure--fragment mass space. The dynamic pressure is the atmospheric 
pressure exerted on a meteoroid or a fragment derived from it just 
before another fragmentation. Results of 7--11 automatic solutions with 
an acceptable quality are shown for each fireball, except for the case 
of EN180118, for which only two solutions of poor quality are shown.}
         \label{pres_mass}
\end{figure}

The results of the automatic and 
manual 
approach can be compared in the dynamic pressure–fragment mass space. It shows the masses of fragments released from the meteoroid and later from 
its fragments, and it also shows whether the same fragmentation times were found by 
both methods, which then leads to the same values of the dynamic pressure. 
Fig.~\ref{pres_mass} shows that both methods give similar 
results (open symbols versus solid symbols of the same shape and 
color). Some differences are also clear: The automatic method 
produces some very low fragment masses, which are allowed by 
the wider span of fragment masses to choose from. The main cause for 
this behavior may be the process of the solution evolution itself. For 
example, the algorithm initially chooses the greater mass of a specific 
fragment, but it later finds that this fragment mass should be 
negligible or zero, which is currently not allowed in the optimization 
function. Therefore, the mass of this fragment evolves to very low 
values until it has little effect on the overall solution (mainly total 
brightness). In a future version of the program, we wish to resolve this 
issue.

This plot can be used to estimate the uncertainties of the fragment masses and 
the dynamic pressure at which the meteoroid and its descendant fragments 
break. A very simplistic solution is to filter all important 
fragments (that contribute to the model light curve) and then calculate 
the arithmetic mean (or median) and a statistical error of the mean of 
their respective masses and dynamic pressures. 

\section{Discussion}\label{discussion}
After modeling the five fireballs presented in this paper, we
describe a few critical points of the modeling. The first point is the 
unequivocal determination of fragmentation times. As we already 
mentioned, gross fragmentations are usually found easily because 
they appear as bright flares in the radiometric curve. This was not 
always the case, however, and then it was more problematic to successfully model 
the fireball fragmentation. The fragmentations that produce eroding 
fragment(s) are even more difficult to find, and it is also not possible 
to set it as precisely as the gross fragmentation. It is usually marked 
by a change in the overall slope of the radiometric curve, but finding such 
a point requires some modeling experience and experimenting. This also 
propagates to the results of the modeling process: it affects 
the derived mechanical strength of the meteoroid or its fragments. 

The second critical point is finding the proper weights for individual 
datasets (radiometric curve, light curve, and dynamic data) when 
calculating the $\chi^2$ sum. This seems to have a substantial effect 
on finding an acceptable solution to fireball fragmentation. We described the 
probable physical interpretation of these weights in detail in Sect.~\ref{dataset-weights} . Nevertheless, it is 
one of the critical values that we need to experimentally assign to 
each run from a plausible range, and we have still little understanding 
of what the value should be beforehand. 

Furthermore, the price for a better fit is usually the production of more 
fragments from the meteoroid and the higher complexity of the model. 
This should mostly be avoided because it is questionable whether the data 
support such a complex model. We recall that the number of free parameters 
grows with the number of fragments. Moreover, we note that the 
automatic model sometimes contains some very little fragments that do 
not manifest themselves in the available data, they are rather 
numerical artifacts. Therefore, we prefer the models with a lower 
number of fragments involved, as in the manual solution, and we follow the principle of 
Occam's razor. We note that this preference may lead to 
biased solutions. In the future version of the program, we will focus 
more on this problem, and we consider that the total number of fragments may 
be one of the criteria for an acceptable solution that can be forced already in the 
optimization process. It should not be difficult to model any data 
with a huge number of free parameters, but it is hardly justifiable.

We do not claim that the current approach can find a unique 
solution to the problem. However, it is mostly able to find a solution 
that is consistent with all the available data we have about 
meteoroid atmospheric fragmentation. We note that as with any numerical 
model, our model is also simplified and cannot capture all the fine 
details of the process. For instance, there could be smaller 
fragments that do not substantially contribute to the overall 
brightness of the fireball, and they do not have an impact on dynamics. 
In this sense, there could be many equally good solutions with a 
different number of fragments for a single fireball, and the 
solution we presented is only one representative solution of this set. 

As noted above, we usually employed a computation cluster comprising 
a few dozen nodes with 48 threads (24 physical cores) each. The 
computation of a single model on a single node took tens of hours to a 
few days, depending on the complexity of optimized data. We will 
consider a few ways to improve the speed of the computation in the 
future.
 
\section{Conclusions}\label{conclusions}
In this paper, we presented the semi-automatic program called FirMpik, which can find quality solutions for fireball fragmentation in the 
atmosphere consistent with the radiometric and photometric data as 
well as the dynamics of the meteoroid.
When compared to the manual approach, the program can find a 
solution of similar quality and sometimes a solution that is formally slightly better. However, we demonstrated on some fireballs that it 
cannot find a substantially better solutions. It also did not find 
solutions that would fit data similarly, but would be markedly 
different. The meteoroid fragments into a similar number of fragments 
that have roughly the same mass; where there is an eroding fragment in 
the manual solution, the automatic one features one (or a few) eroding 
fragments as well. This was shown in Sect.~\ref{results}. 

These automatic solutions consistently map in the same region of the 
dynamic pressure–fragment mass space as the manual solutions. The only 
difference is that they often comprise a greater total number of 
fragments than the manual solutions. In the current version of the 
program, the number of fragments for each fragmentation time is 
selected randomly from a hard-wired span of values, but more constraints can be placed on this process in the future version of the program.

The uncertainties of the calculated quantities (dynamic pressure, fragment 
masses, initial meteoroid velocity, and mass) were estimated from the 
spread of the values in different solutions of similar quality. This was
not possible before in the manual procedure, where only one solution was 
found for each fireball. However, we were not able to find a more 
rigorous and computationally viable way to calculate uncertainties.

Currently, we have to find the fragmentation times in the data manually 
and then increase their precision with an automatic procedure. Then, we 
estimate a plausible range of weights of available datasets and run 
several (tens of) simulations to find the most plausible value. At the 
same time, we employ a wide range of the initial meteoroid mass and 
velocity values and a wider range of fragment numbers in each 
fragmentation time. Later, when we find more viable types of solutions, 
we narrow these ranges down, which enables a faster search of the 
parametric space, and we find a global optimum of the problem more likely.

We usually need to run a few to several tens of runs before we find 
a quality fit to the fireball data. This is necessary because of the 
inherent stochasticity of the optimization process. During the 
computation, the algorithm extensively uses pseudo-random numbers, and 
it is also randomly initialized. Still, the algorithm converges to 
similar solutions that have a comparable $\chi^2$ value, as we expected. 
The nature of the GA also promotes a thorough search of 
the parametric space, because several tens to hundreds of 
individuals are dispersed and search independently for the solution. 
Moreover, the data we have about meteoroid fragmentation are 
complementary and further constrain the potential solution. Therefore, 
 the algorithm is in most cases capable of finding a 
global optimum that fits all the available data. With all the model 
simplifications, this solution consistently describes the fireball 
fragmentation in the atmosphere.

In the future, we would like to improve finding the fragmentation times, 
make it more robust, and make it part of the optimization process. A more in-depth understanding of the dataset weights is our next goal as well. 
Last, we would like to place more constraints on the model to be more 
confident about the uniqueness of the solution by employing more data 
for this procedure, for instance, data on individual fragments observed for some 
fireballs with narrow-field-of-view cameras.

\begin{acknowledgements}
We are grateful to Donovan Mathias for a very careful and constructive 
review that helped us substantially improve the presentation of our 
research. This research was supported by grant no. 19-26232X from the 
Czech Science Foundation. The computations were performed on the OASA 
and VIRGO clusters of the Astronomical Institute of the Czech Academy 
of Sciences. This research has made use of NASA’s Astrophysics Data 
System.
\end{acknowledgements}

\bibliographystyle{aa}
\bibliography{aa45023}

\begin{appendix}
\section{Additional figures}\label{app_fig}
In this appendix, we present additional figures that compare the results 
of the automatic and the manual modeling of the other four bolides. They are described in detail in Sect.~\ref{results}.

\begin{figure}[!h]
   \centering
   \includegraphics[width=\hsize]{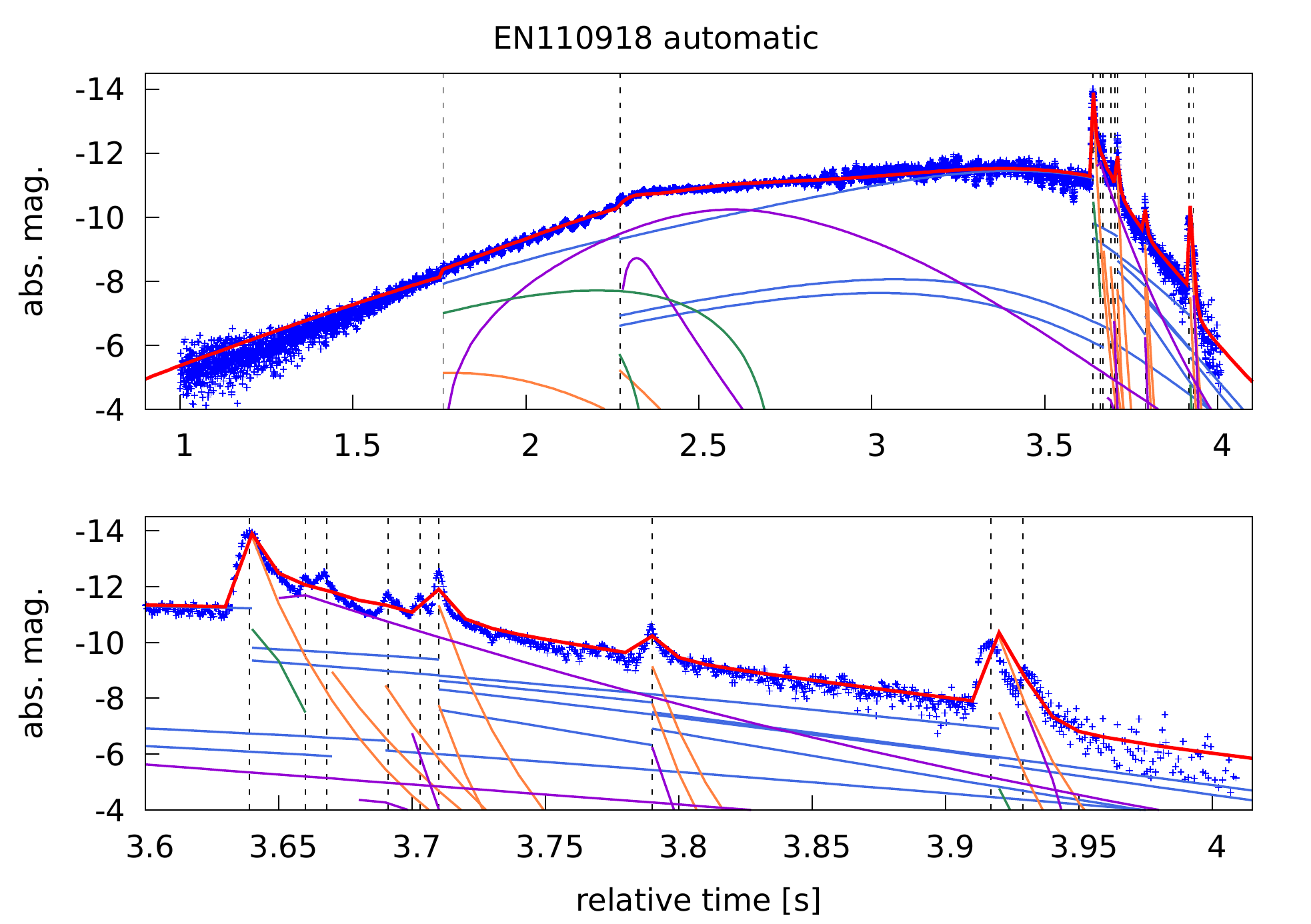}
      \caption{Automatic solution for EN110918 in a lower time 
resolution of $t_{\rm res}=0.01\,\rm{s}$. The upper panel shows the 
whole radiometric curve, and the lower panel shows the end of it in 
greater detail. Some of the details at the end of the radiometric curve 
are not present in the model. Labels are the same as in 
Fig.~\ref{260815_best_rlc}.}
         \label{110918_low_timeres_rlc}
\end{figure}

\begin{figure}[!h]
   \centering
   \includegraphics[width=\hsize]{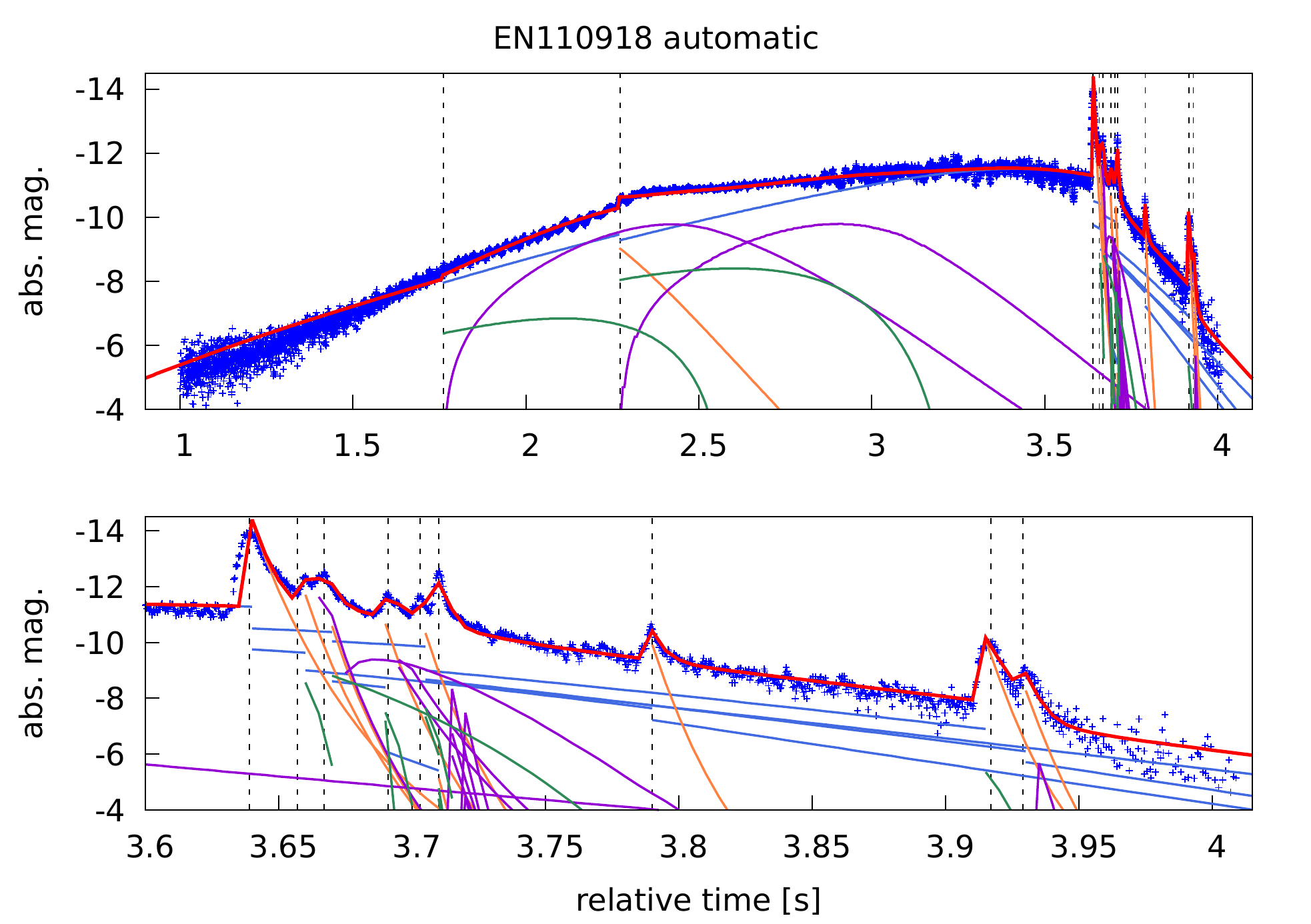}
      \caption{Best obtained automatic solution to EN110918 in 
higher time resolution of $t_{\rm res}=0.005\,\rm{s}$. Labels are the same as 
in Fig.~\ref{260815_best_rlc}.}
         \label{110918_best_rlc}
\end{figure}

\begin{figure}
   \centering
   \includegraphics[width=\hsize]{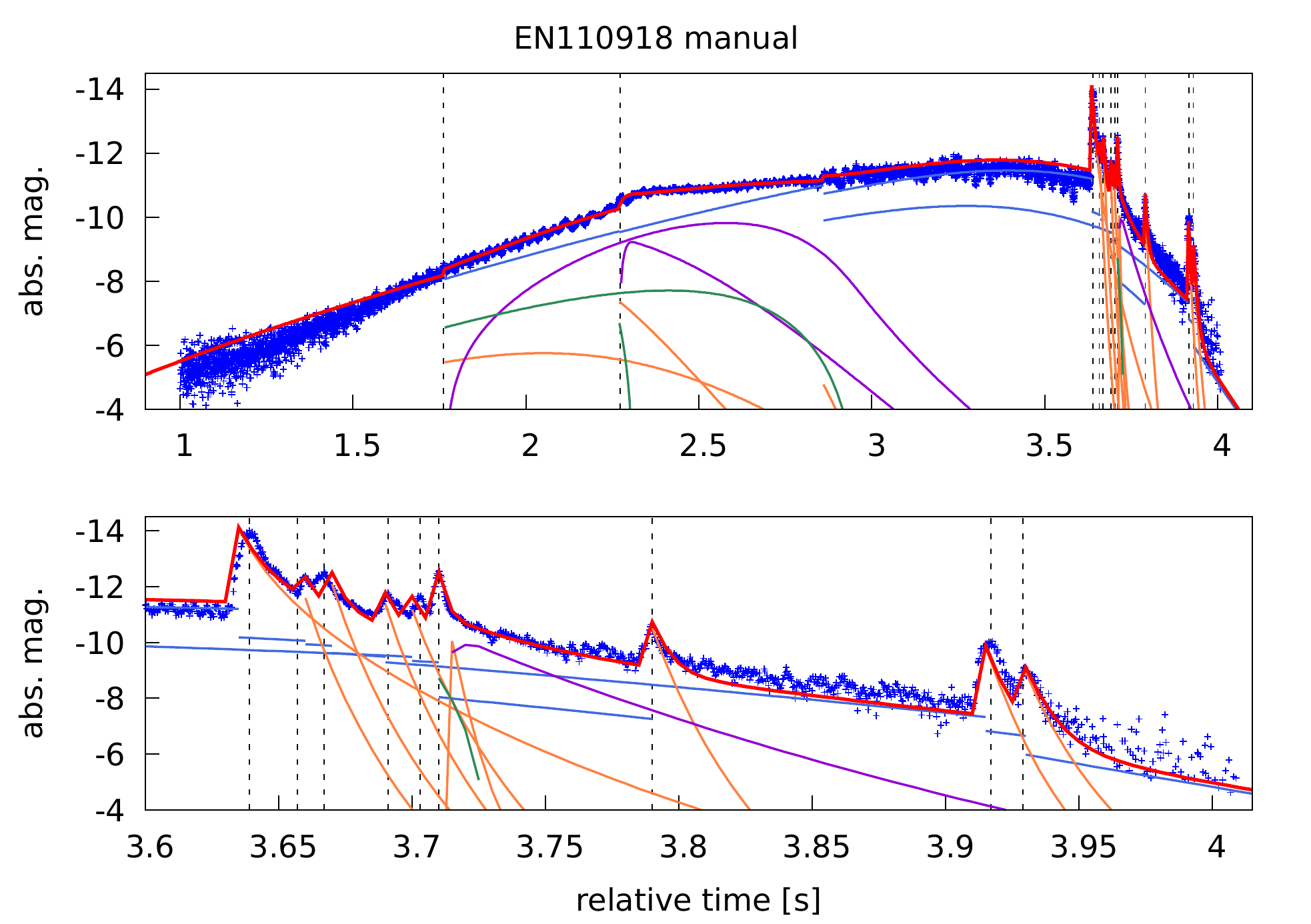}
      \caption{Manual solution for EN110918 with the same time 
resolution as the best automatic solution. Labels are the same as in 
Fig.~\ref{260815_best_rlc}.}
         \label{110918_man_rlc}
\end{figure}

\begin{figure}
   \centering
   \includegraphics[width=\hsize]{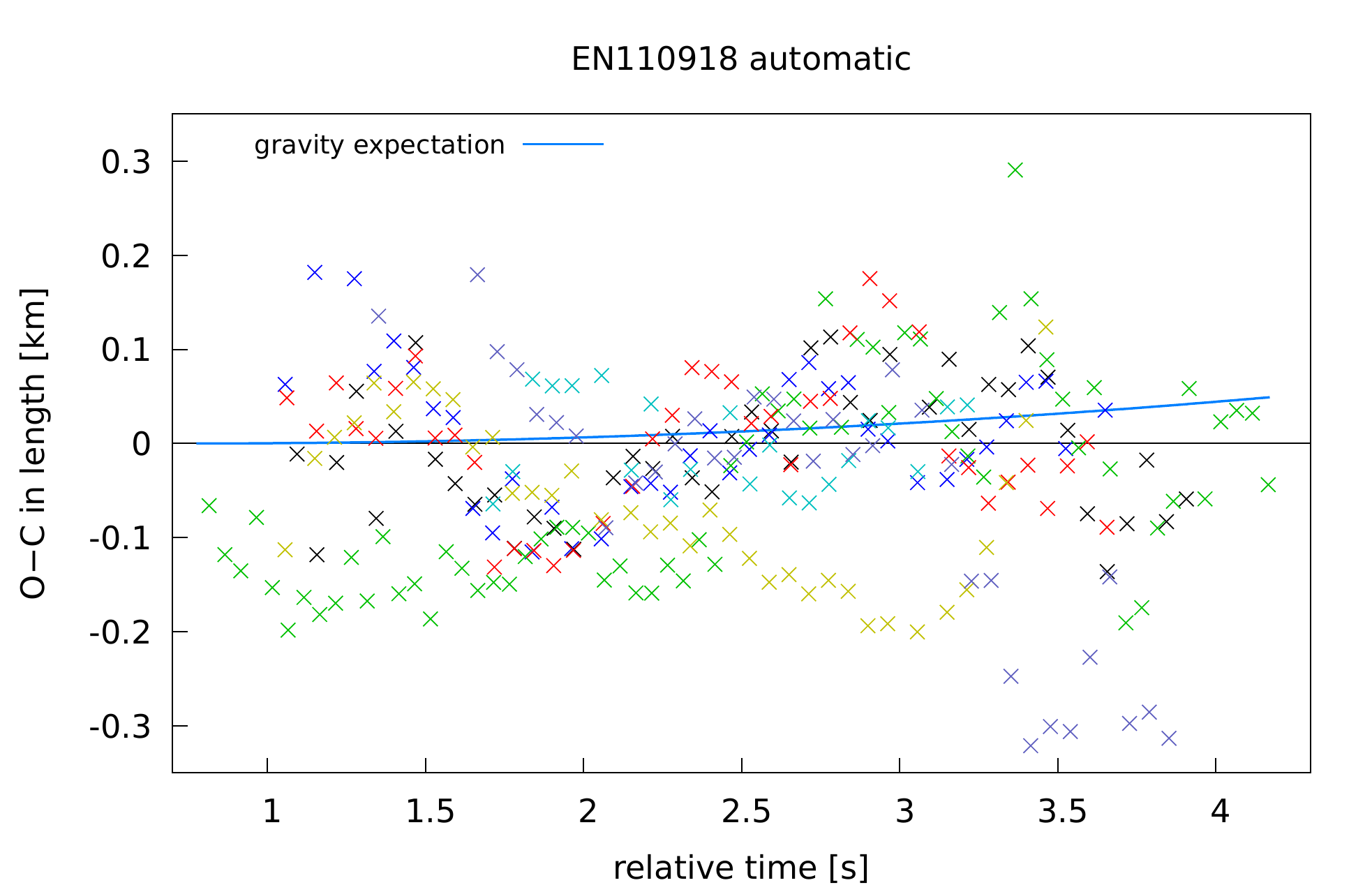}
      \caption{Length residuals of the automatic model of EN110918. 
Labels are the same as in Fig.~\ref{260815_best_len}.}
         \label{110918_best_len}
\end{figure}

\begin{figure}
   \centering
   \includegraphics[width=\hsize]{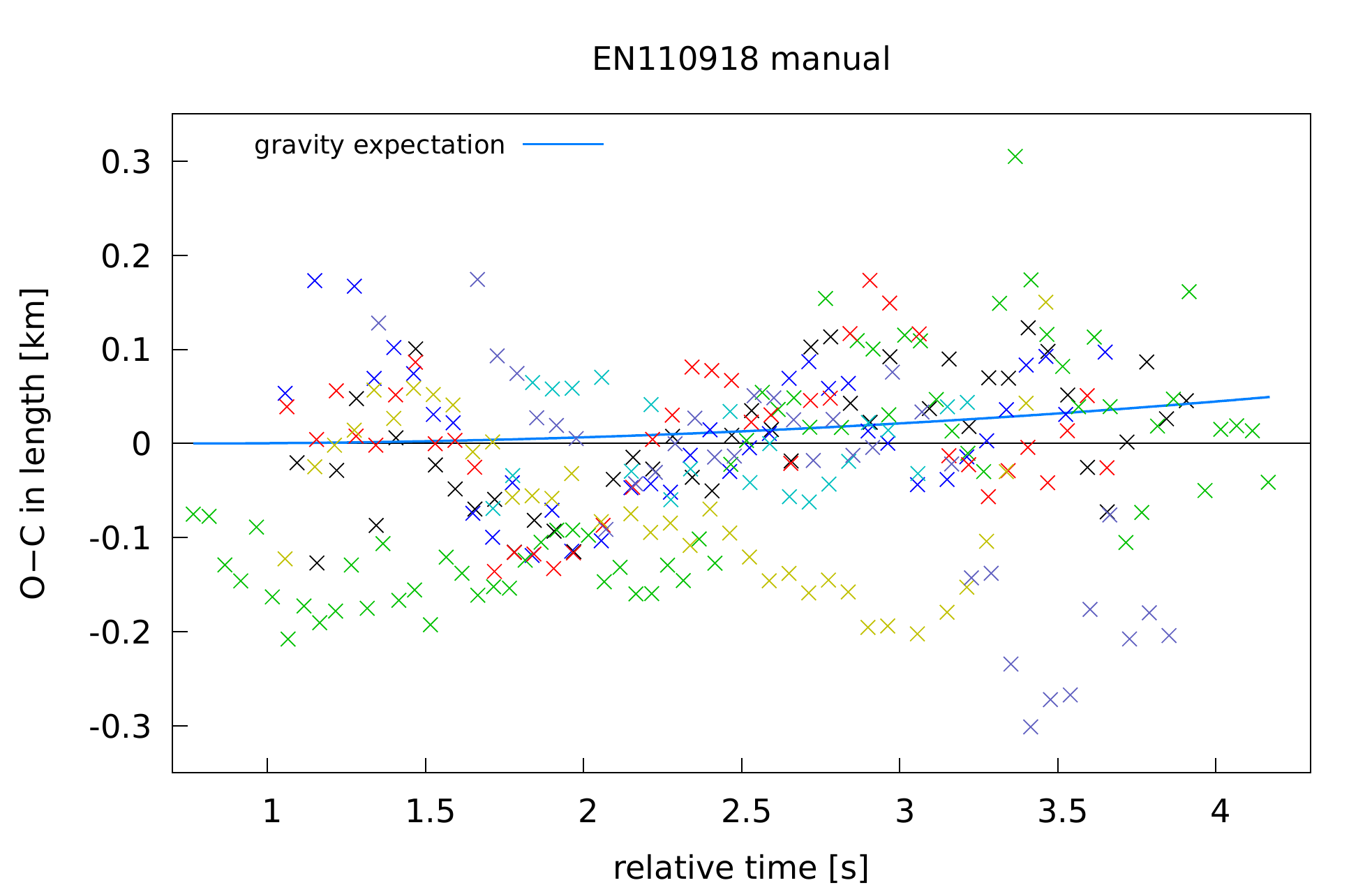}
      \caption{Length residuals of the manual model of EN110918. 
Labels are the same as in Fig.~\ref{260815_best_len}.}
         \label{110918_man_len}
\end{figure}

\begin{figure}
   \centering
   \includegraphics[width=\hsize]{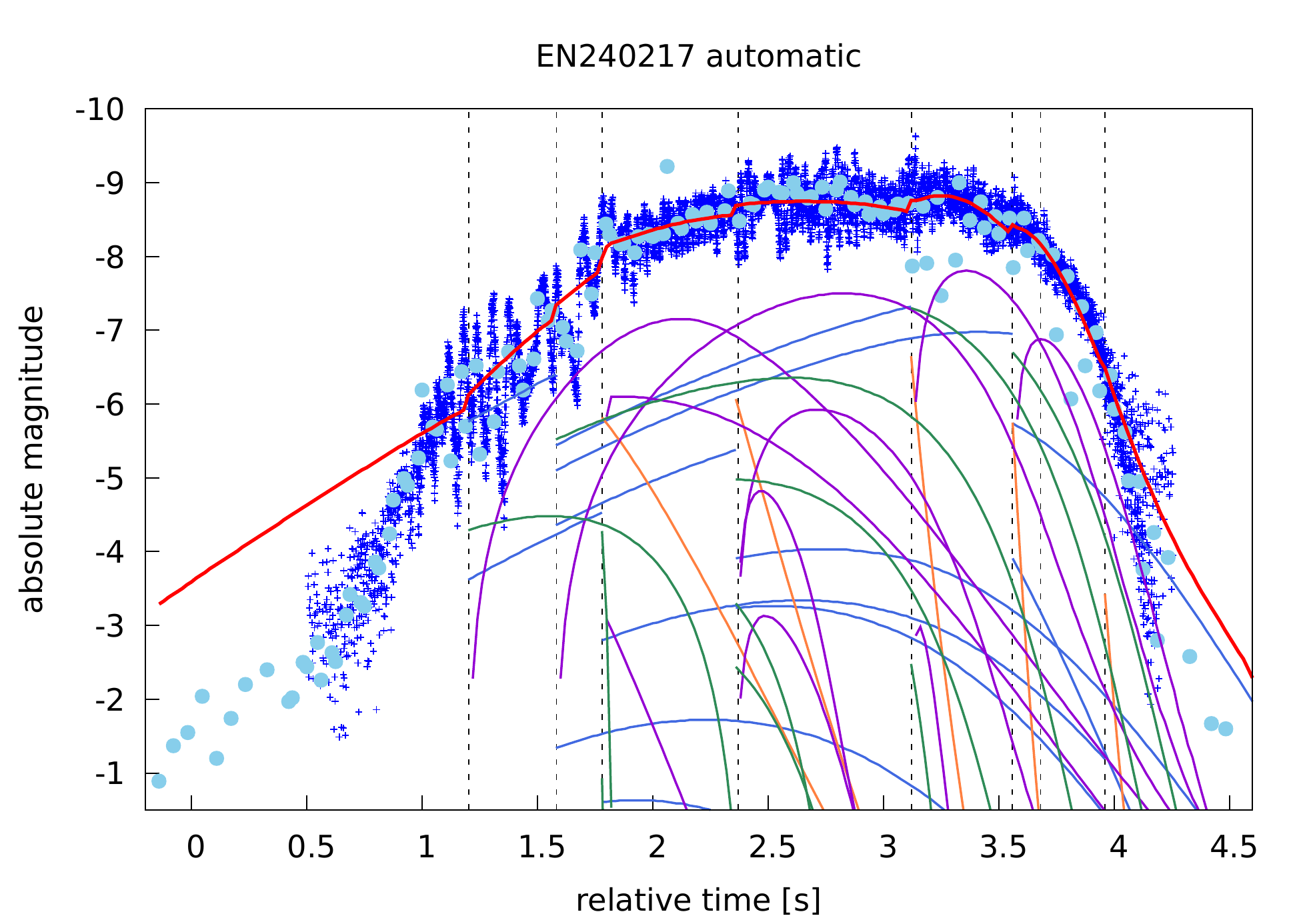}
      \caption{Automatic solution of EN240217 compared to 
radiometric and light-curve data. Labels are the same as in 
Fig.~\ref{260815_best_rlc}.}
         \label{240217_best_rlc}
\end{figure}

\begin{figure}
   \centering
   \includegraphics[width=\hsize]{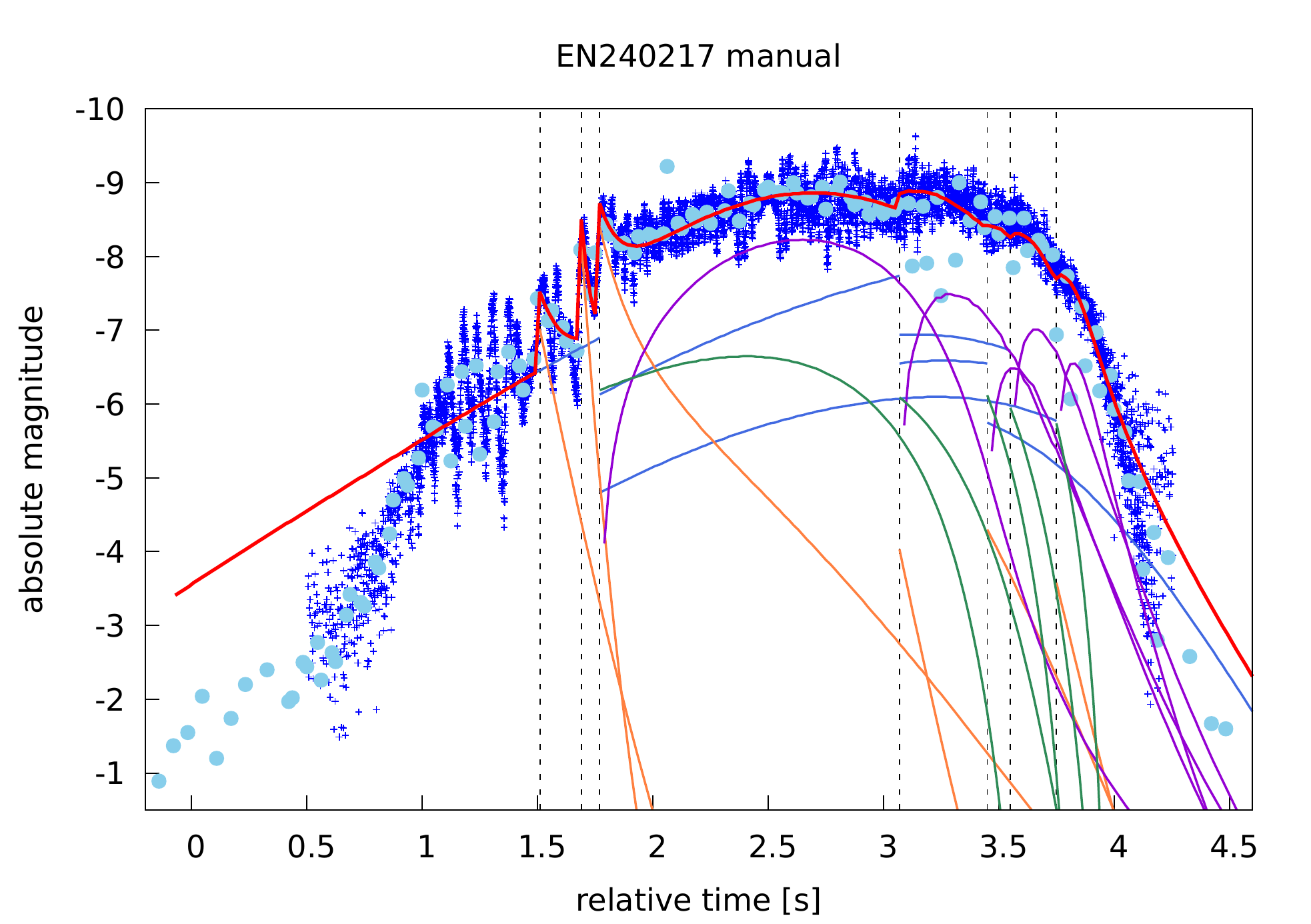}
      \caption{Manual solution for EN240217. Labels are the same as in 
Fig.~\ref{260815_best_rlc}.}
         \label{240217_man_rlc}
\end{figure}

\begin{figure}
   \centering
   \includegraphics[width=\hsize]{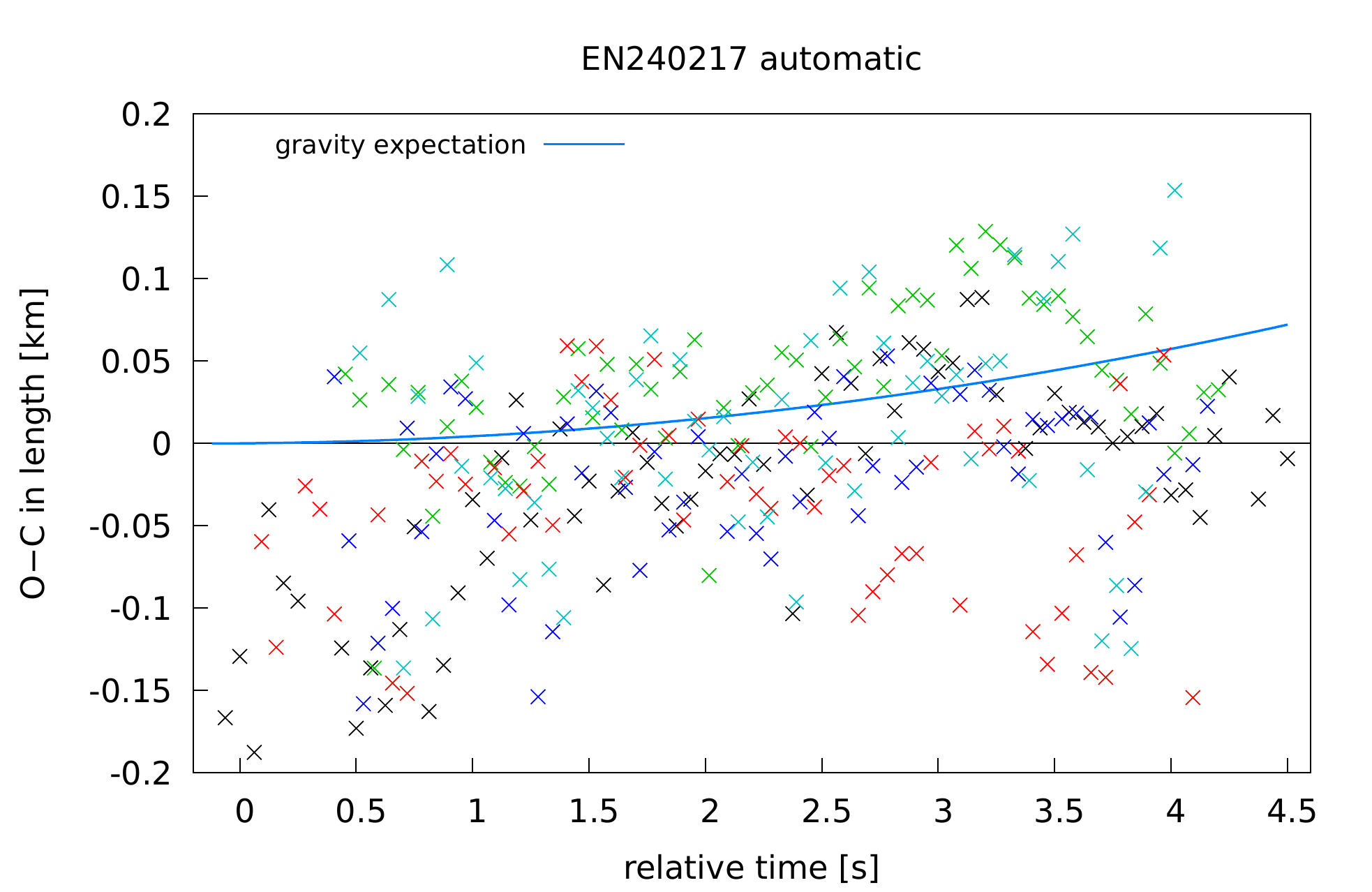}
      \caption{Length residuals of the automatic model for EN240217. 
Labels are the same as in Fig.~\ref{260815_best_len}.}
         \label{240217_best_len}
\end{figure}

\begin{figure}
   \centering
   \includegraphics[width=\hsize]{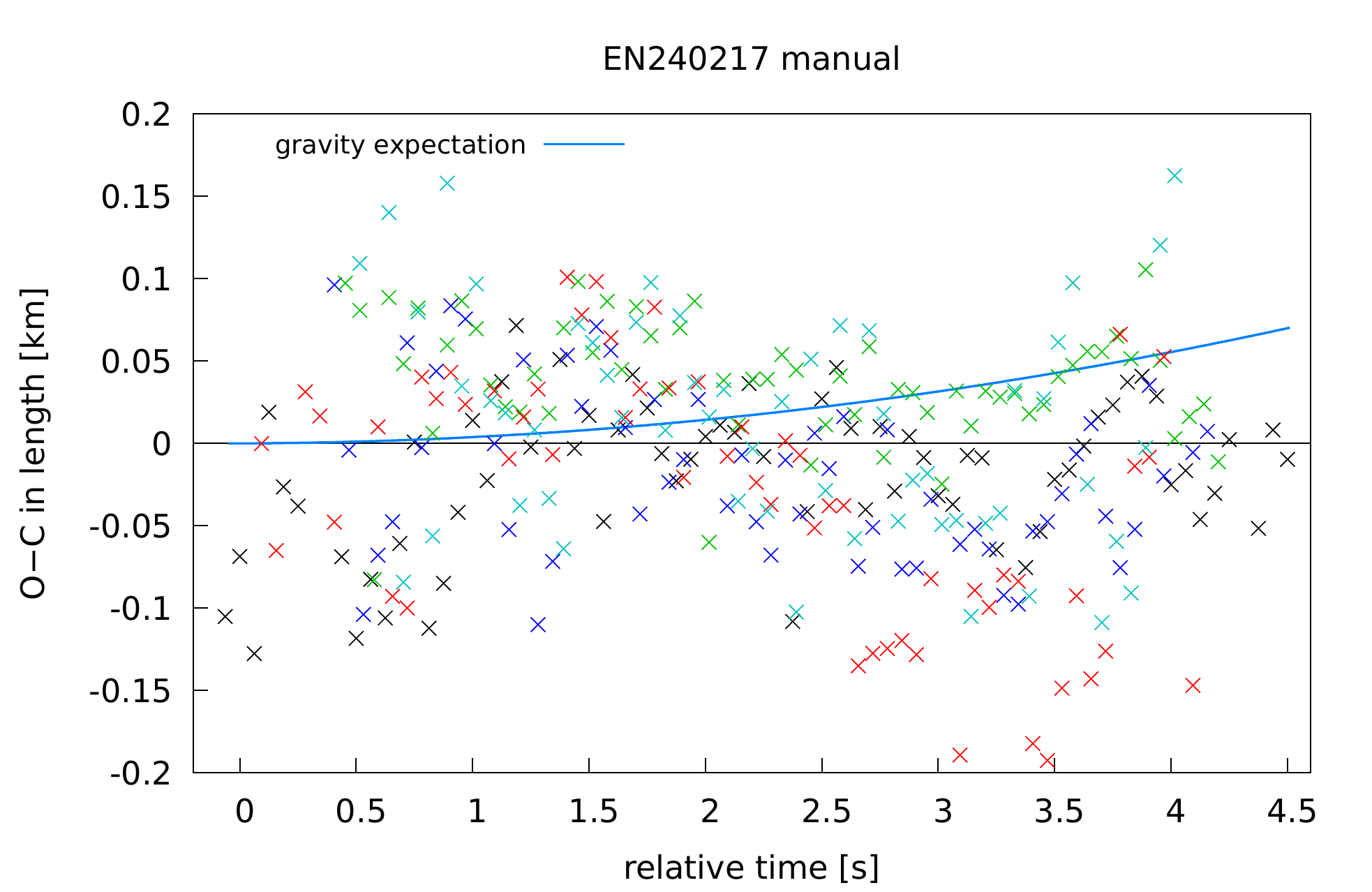}
      \caption{Length residuals of the manual model for EN240217. 
Labels are the same as in Fig.~\ref{260815_best_len}.}
         \label{240217_man_len}
\end{figure}

\begin{figure}
   \centering
   \includegraphics[width=\hsize]{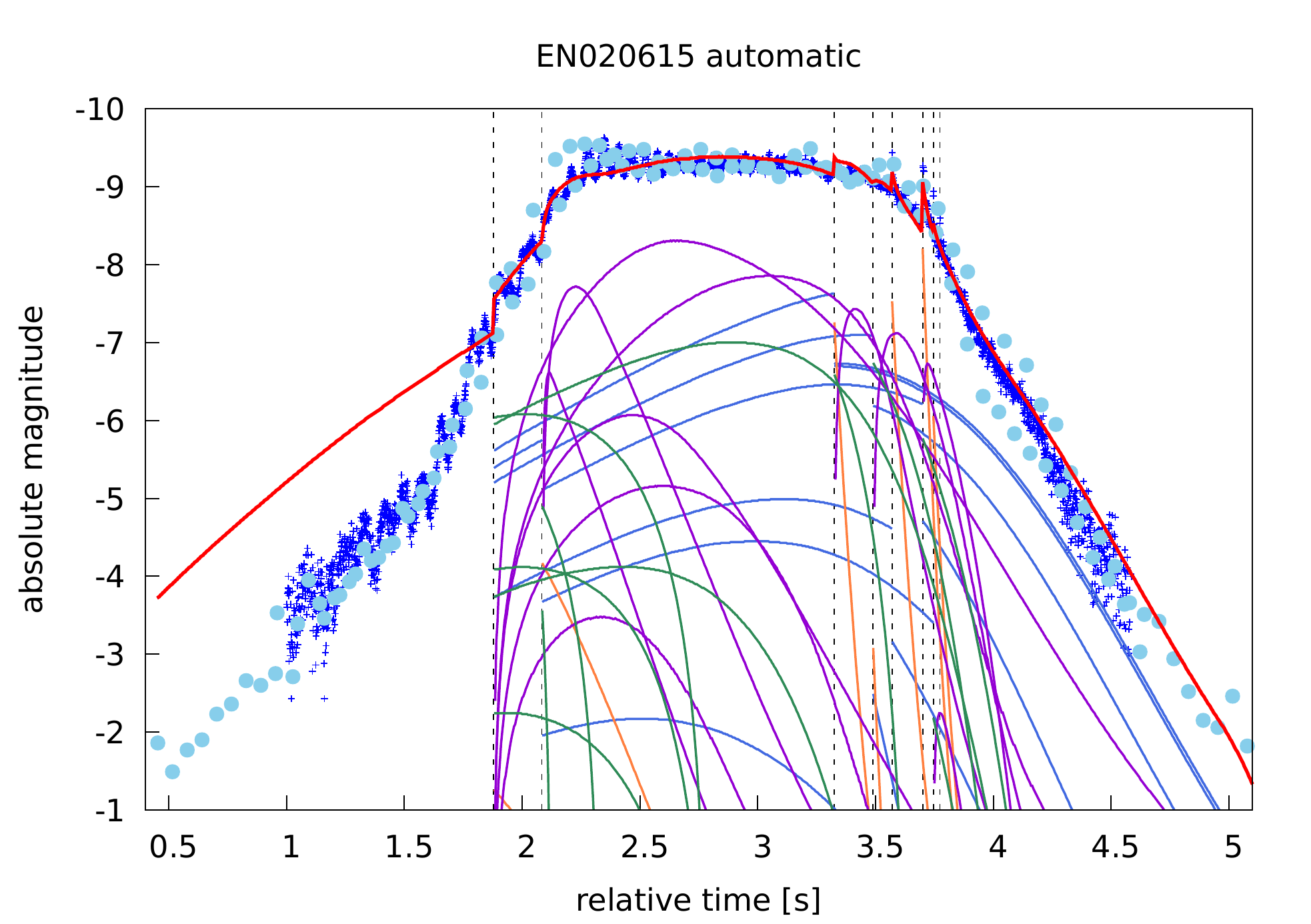}
      \caption{Automatic solution for EN020615 compared to 
radiometric and light-curve data. Labels are the same as in 
Fig.~\ref{260815_best_rlc}.}
         \label{020615_best_rlc}
\end{figure}

\begin{figure}
   \centering
   \includegraphics[width=\hsize]{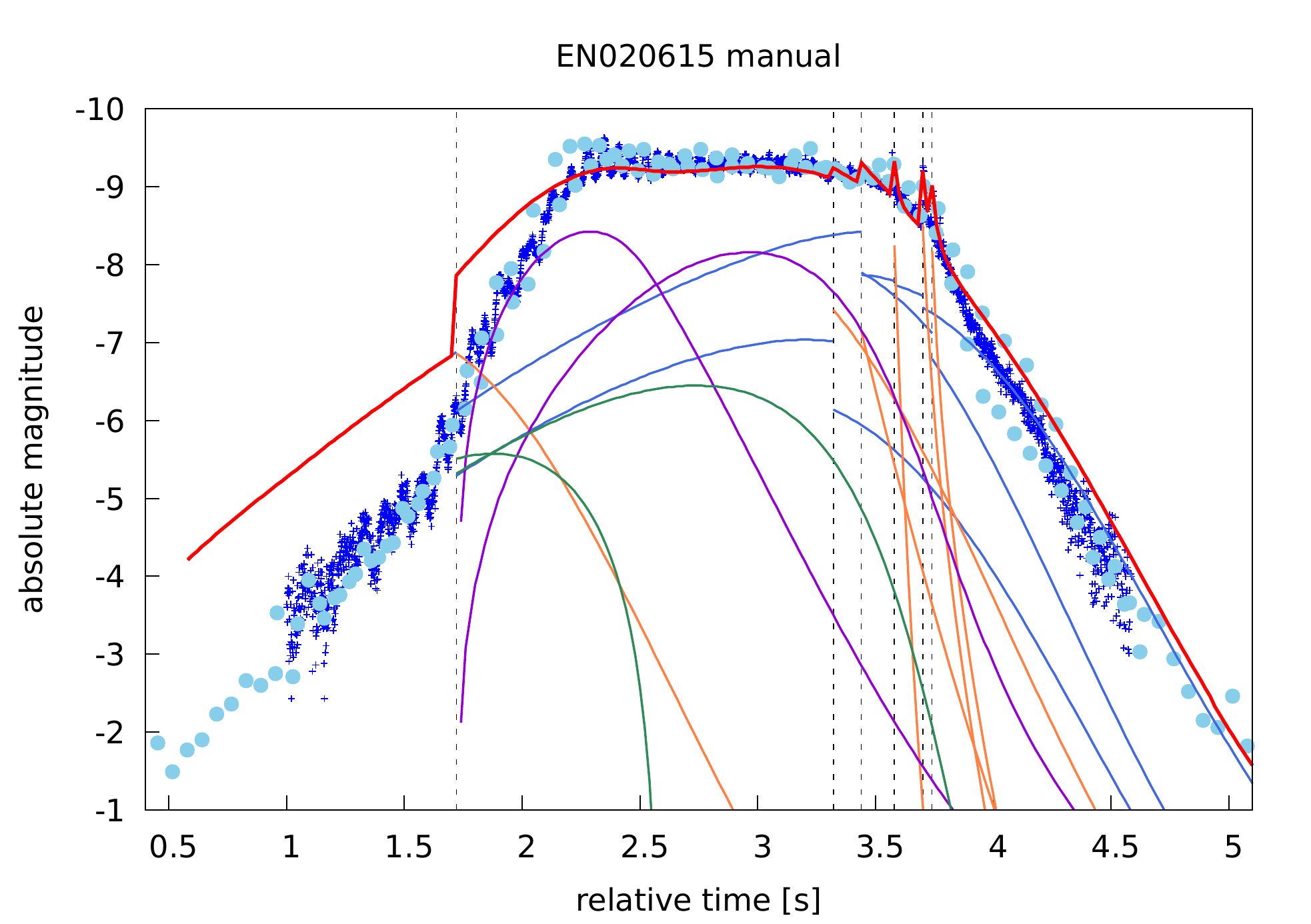}
      \caption{Manual solution for EN020615. Labels are the same as in 
Fig.~\ref{260815_best_rlc}.}
         \label{020615_man_rlc}
\end{figure}

\begin{figure}
   \centering
   \includegraphics[width=\hsize]{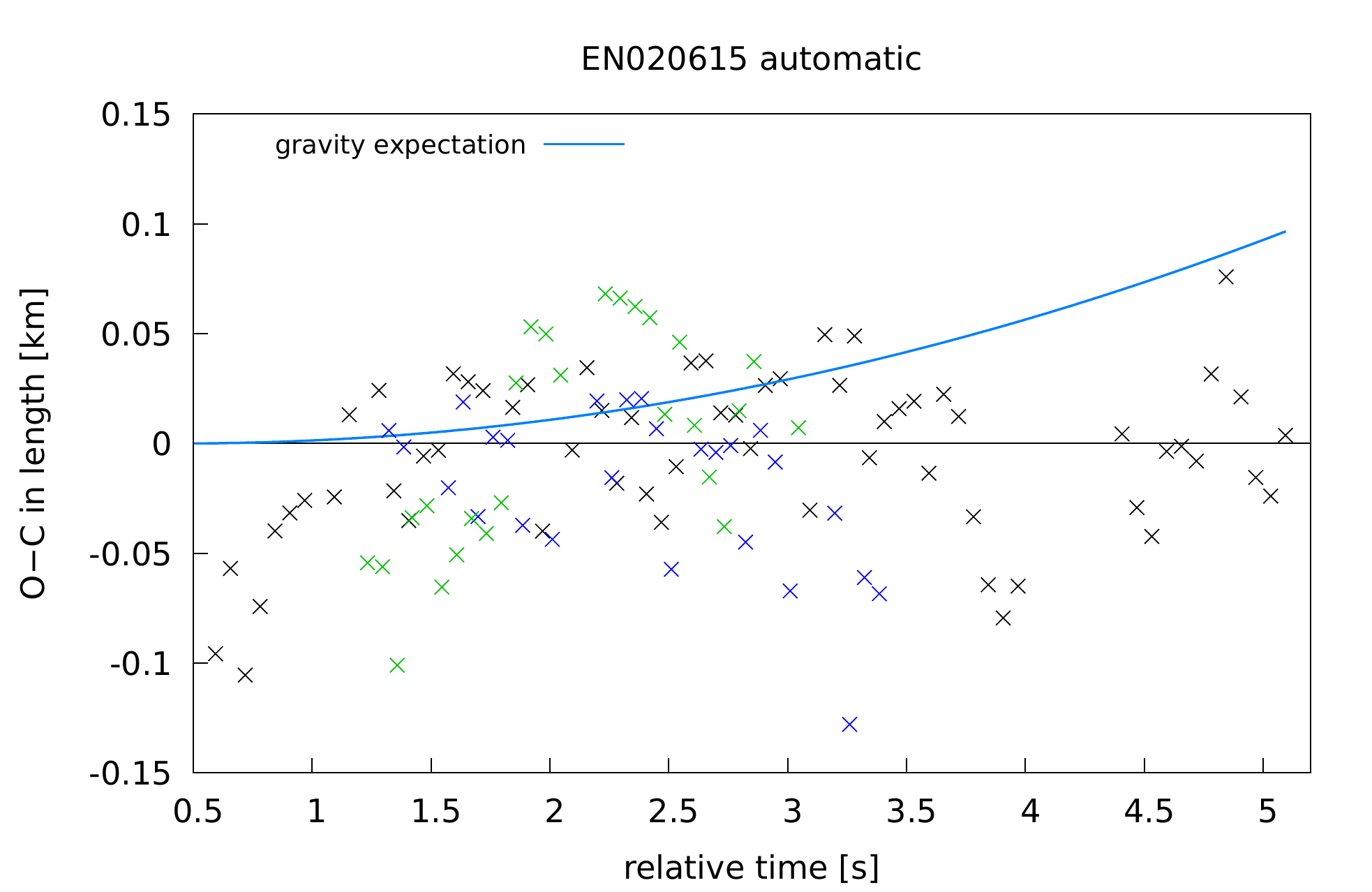}
      \caption{Length residuals of the automatic model for EN020615. 
Labels are the same as in Fig.~\ref{260815_best_len}.}
         \label{020615_best_len}
\end{figure}

\begin{figure}
   \centering
   \includegraphics[width=\hsize]{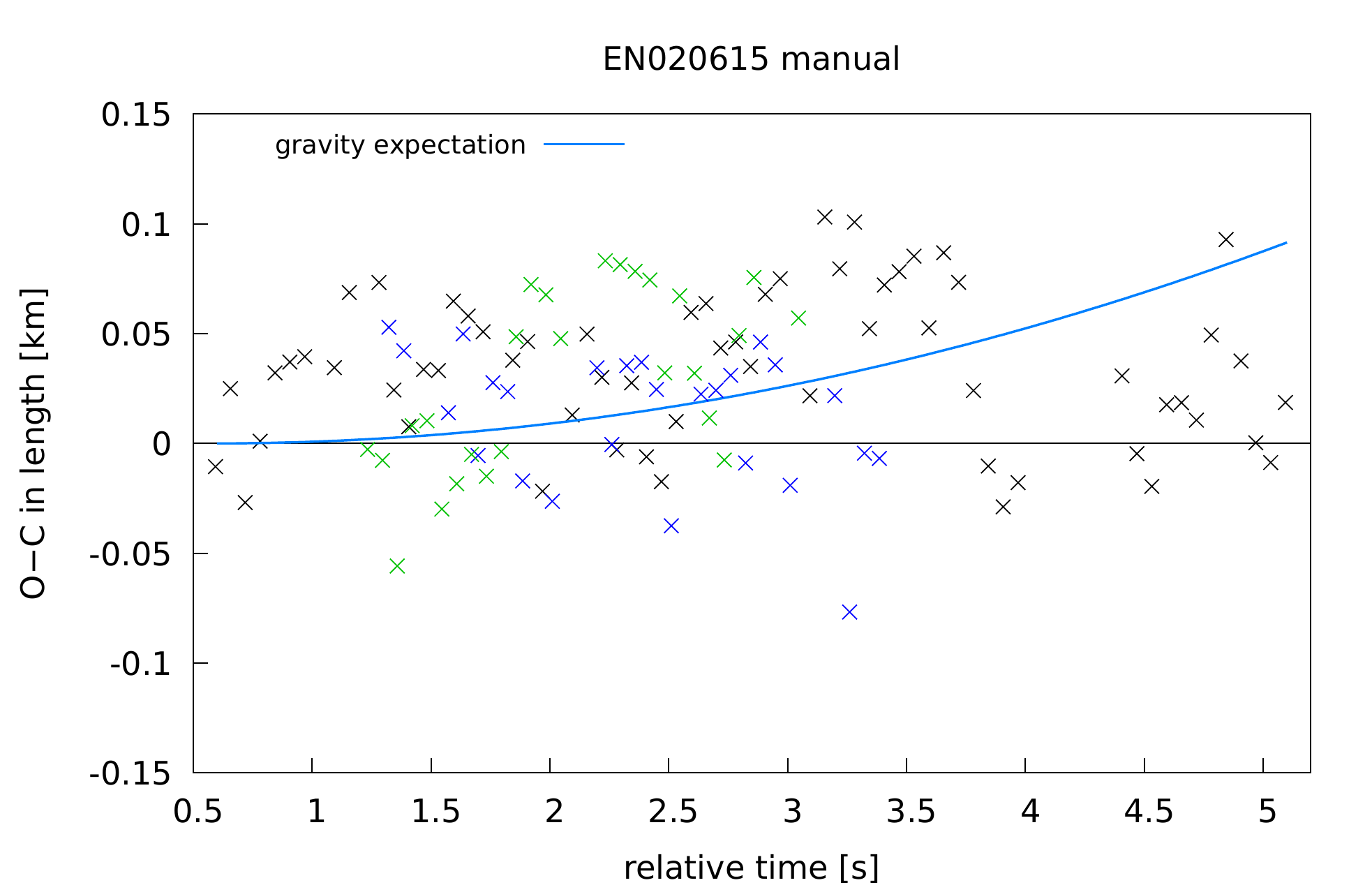}
      \caption{Length residuals of the manual model for EN020615. 
Labels are the same as in Fig.~\ref{260815_best_len}.}
         \label{020615_man_len}
\end{figure}

\begin{figure}
   \centering
   \includegraphics[width=\hsize]{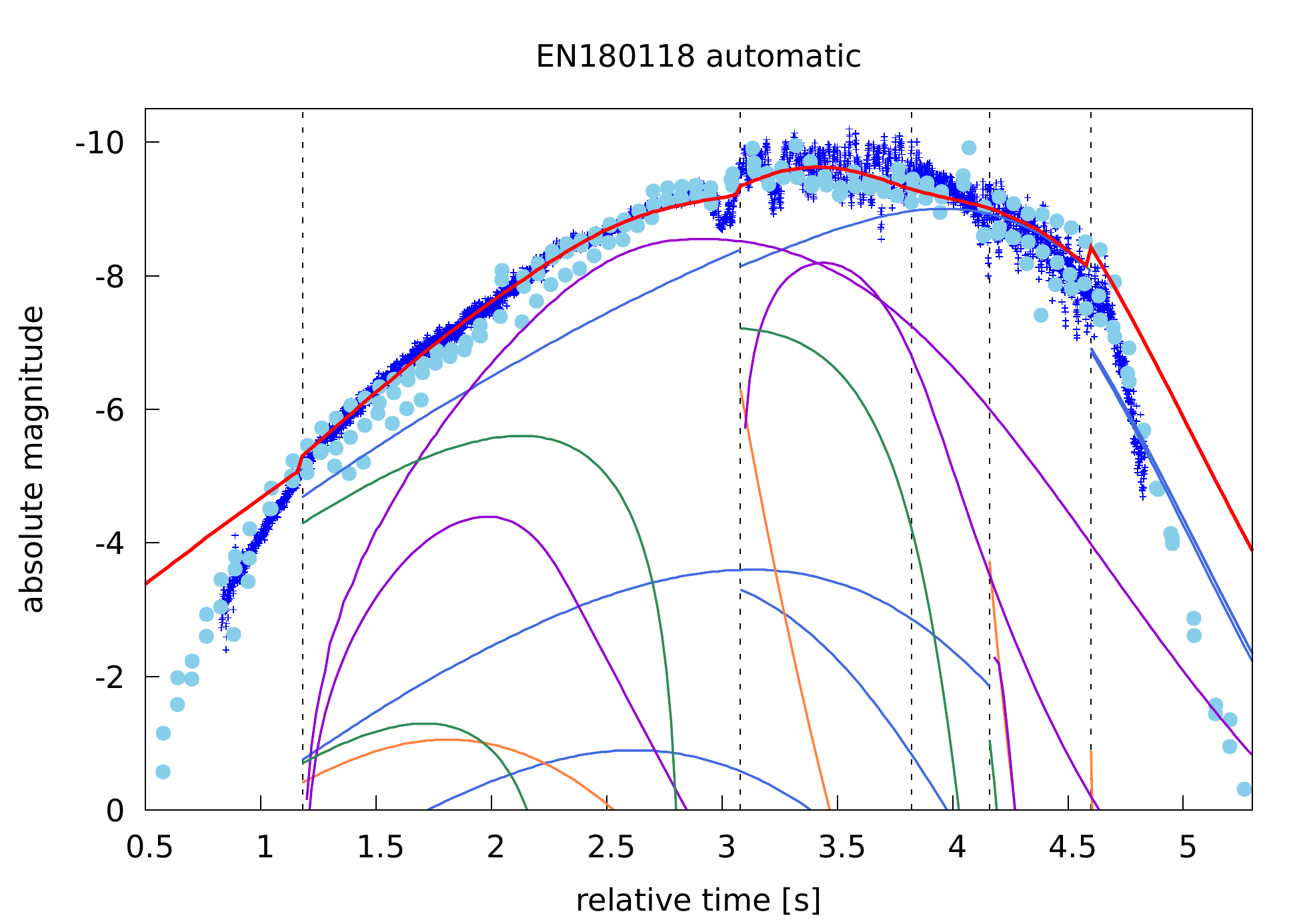}
      \caption{Automatic solution for EN180118 compared to 
radiometric and light-curve data. Labels are the same as in 
Fig.~\ref{260815_best_rlc}.}
         \label{180118_best_rlc}
\end{figure}

\begin{figure}
   \centering
   \includegraphics[width=\hsize]{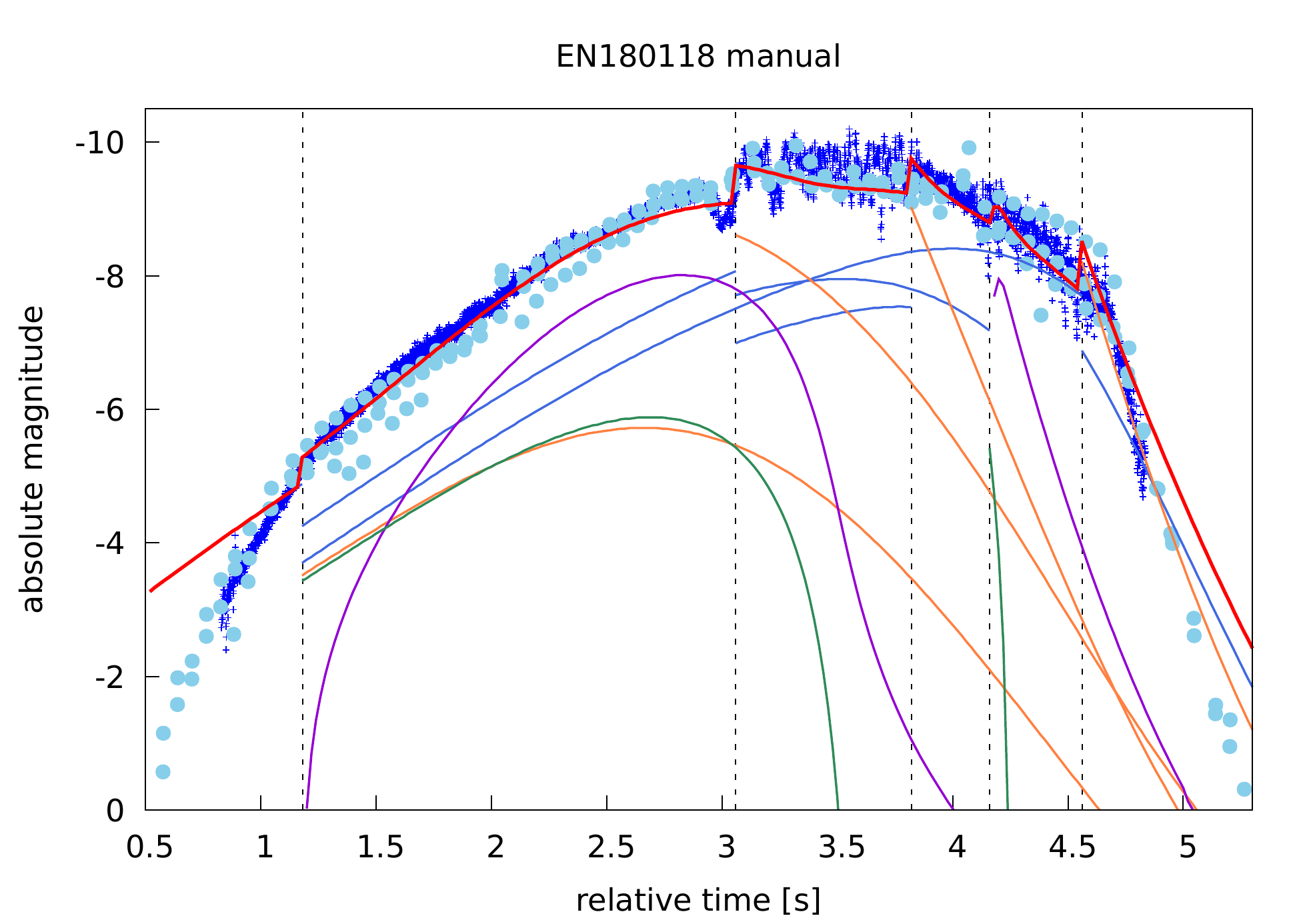}
      \caption{Manual solution for EN180118. Labels are the same as in 
Fig.~\ref{260815_best_rlc}.}
         \label{180118_man_rlc}
\end{figure}

\begin{figure}
   \centering
   \includegraphics[width=\hsize]{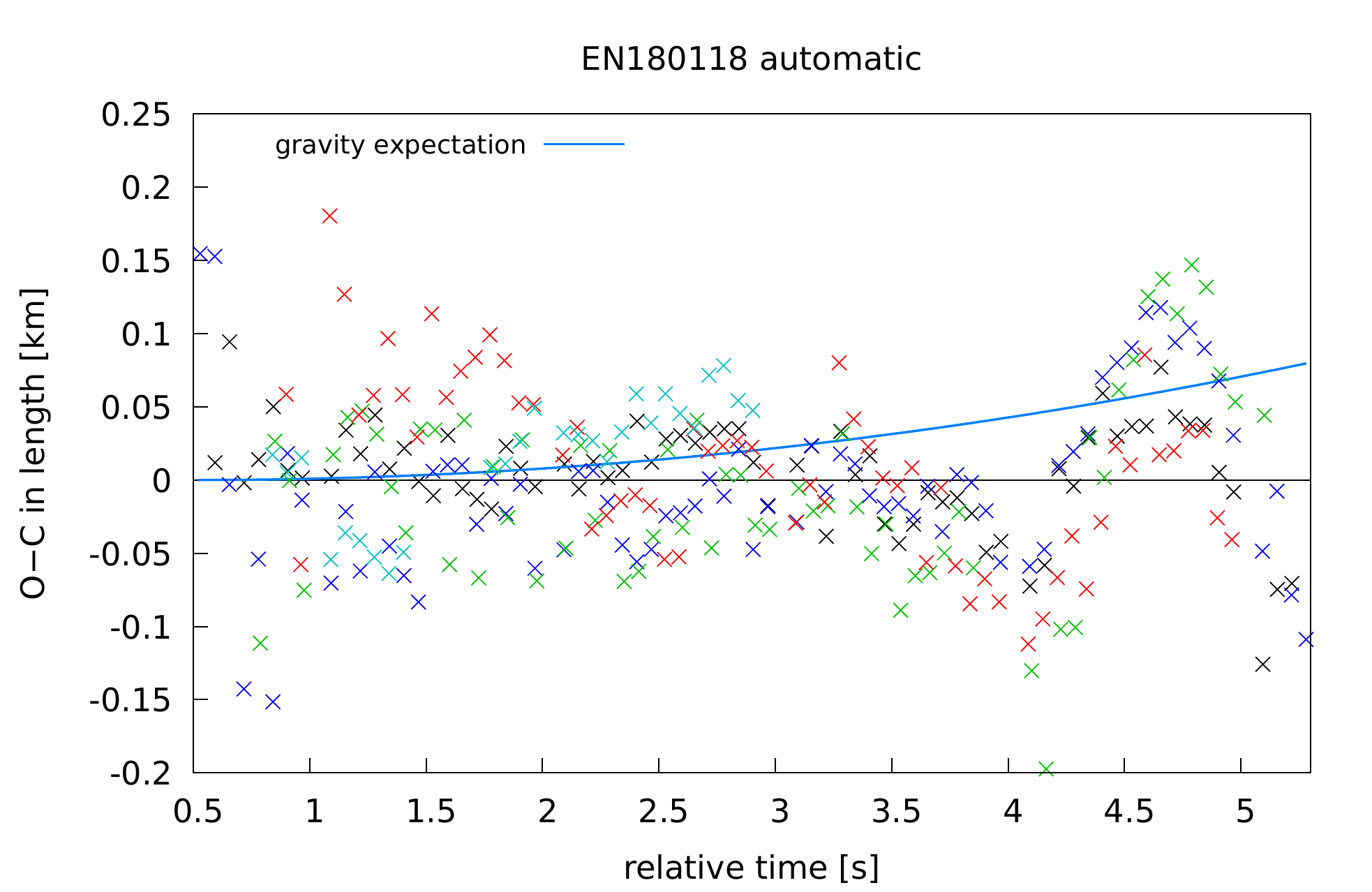}
      \caption{Length residuals of the automatic model for EN180118. 
Labels are the same as in Fig.~\ref{260815_best_len}.} 
         \label{180118_best_len}
\end{figure}

\begin{figure}
   \centering
   \includegraphics[width=\hsize]{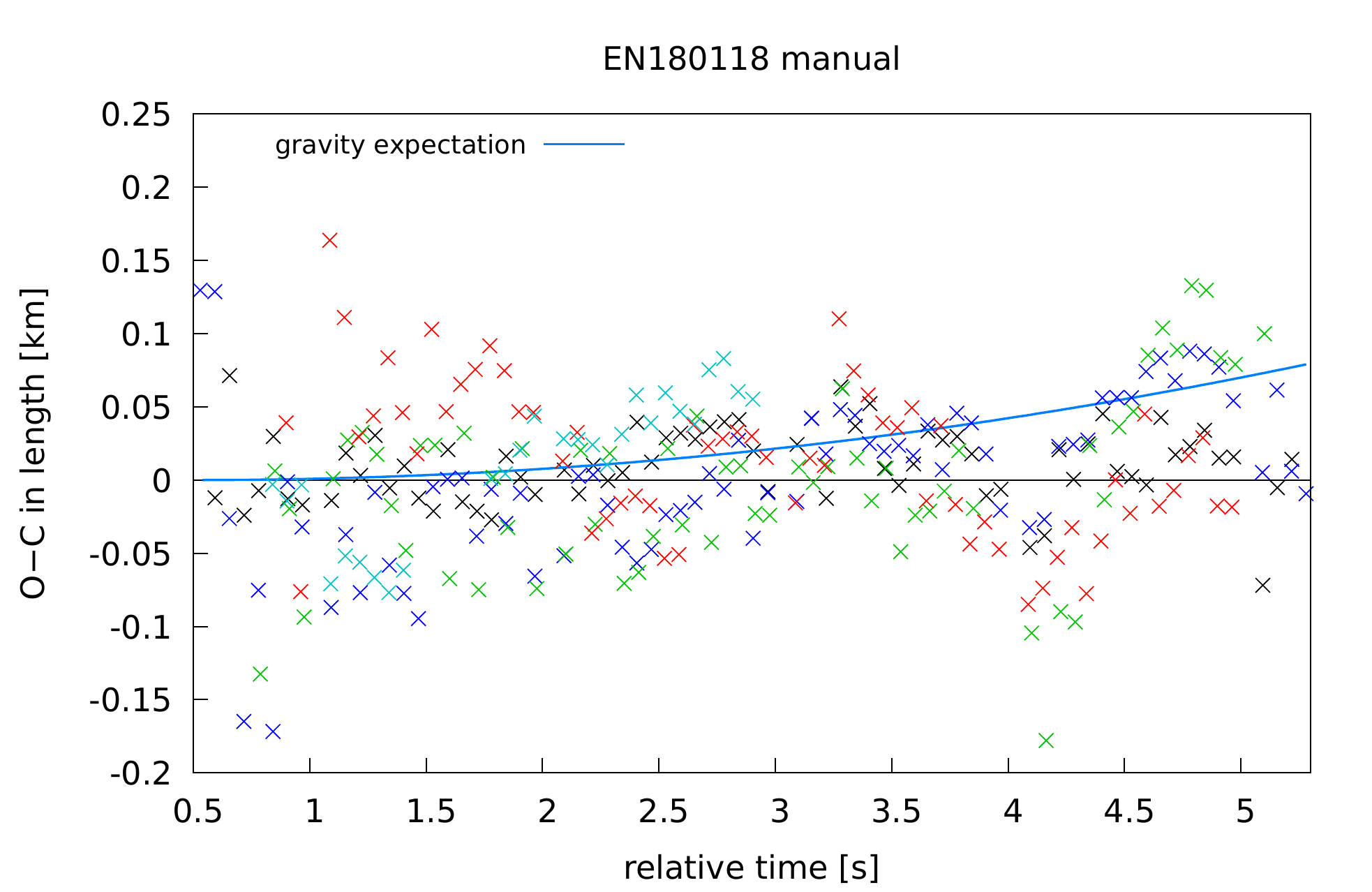}
      \caption{Length residuals of the manual model for EN180118. 
Labels are the same as in Fig.~\ref{260815_best_len}.}
         \label{180118_man_len}
\end{figure}

\end{appendix}
\end{document}